\documentclass[namedreferences]{SolarPhysics}
\usepackage[optionalrh]{spr-sola-addons} % For Solar Physics

\usepackage{graphicx}        % For eps figures, newer & more powerfull
\usepackage{color}           % For color text: \color command
\usepackage{url}             % For breaking URLs easily trough lines
\newcommand{\aap}{    {\it Astron. Astrophys.}}

\newcommand{\apj}{    {\it Astrophys. J.}}

\newcommand{\jgr}{    {\it J. Geophys. Res.}}

\newcommand{\solphys}{{\it Solar Phys.}}

\newcommand{\ssr}{    {\it Space Sci. Rev.}}

\begin{document}

\begin{article}

\begin{opening}

\title{The beginning of halo coronal mass ejection\\ {\it Solar Physics}}

\author{V.G.~\surname{Fainshtein}$^{1}$\sep
        Yu.S.~\surname{Zagaynova}\sep
       }
\runningauthor{Fainshtein and Zagaynova} \runningtitle{The kinematics properties of halo coronal mass ejection}

   \institute{$^{1}$ Institute of Solar Terrestrial Physics Russian Academy of Sciences, Siberian Branch,Irkutsk, Russia
                     email: \url{vfain@iszf.irk.ru}\\
             }

\begin{abstract}
From the GOES-12/SXI data, we studied the initial stage of motion for six rapid (over 1500 km/s) "halo" coronal mass ejections (HCMEs) and traced the motion of these HCMEs within the SOHO/LASCO C2 and C3 field-of-view. For these HCMEs the time-dependent location, velocity and acceleration of their fronts were revealed. The conclusion was drawn that two types of CME exist depending on their velocity time profile. This profile depends on the properties of the active region where the ejection emerged. CMEs with equal ejection velocity time dependence originate form in the same active region. All the HCMEs studied represent loop-like structures either from the first moment of recording or a few minutes later. All the HCMEs under consideration start their translational motion prior to the associated X-ray flare onset. The main acceleration time (time to reach the highest velocity within the LASCO/C2 field-of-view) is close to the associated flare X-ray radiation intensity rise time. The results of (Zhang and Dere, 2006) on the existence of an inverse correlation between the acceleration amplitude and duration, and also on the equality of the measured HCME main acceleration duration and the associated flare soft X-ray intensity rise time are validated. We established some regularities in the temporal variation of the angular size, trajectory, front width and the HCME longitude-to-cross size ratio.
\end{abstract}
\end{opening}
%-------------------------------------------------

\section{Introduction}
     \label{S-Introduction}

Examining the properties of a coronal mass ejection (CME) at the initial stage of its motion is a necessary stage in the process of identifying the physical mechanisms involved in CME formation. Over the last 10 years, a considerable amount of work has been carried out in which CME kinematics was studied right after their formation using the data of various telescopes (see Gallagher et al., 2003; Zhang and Dere, 2006; Temmer el al., 2008; Maričić et al., 2009; Patsourakos et al., 2010; Temmer et al., 2010 and the quoted literature in these papers). These papers obtained important results regarding the initial stage of CME motion.

  %{\S}{\bf --- main conclusions of early/others authors} \\
Zhang and Dere (Zhang and Dere, 2006) came to the conclusion that CME motion can be subdivided into three stages: \textit{i\textup{)}} initial stage, when CME speed slowly increases, \textit{ii\textup{)}} main acceleration phase lasting several minutes to several hours, and \textit{iii\textup{)}} quiet motion phase at an approximately constant speed. They showed that the post-acceleration CME speed and kinetic energy correlate with the maximum value of soft x-ray radiation intensity $I_{\rm SXR}(t)$ from  the flare area relating to the CME (Moon et al., 2003; Burkepile et al., 2004; Vr\v{s}nak et al., 2005). The flare and CME are considered to be related if the place and moment of their emergence are close. It has been established that the kinematics of many CMEs is synchronised with the $I_{\rm SXR}(t)$ behaviour (Gallagher et al., 2003; Zhang and Dere, 2006; Mari\v{c}i\'{c}et al., 2007; Patsourakos et al., 2010). This manifests itself in the time profile of CME speed $V(t)$ in its main acceleration phase being close to the time dependence of the soft X-ray radiation (SXR) intensity, $I_{\rm SXR}(t)$, from the CME-related flare area. There is an inverse correlation between the main acceleration of a CME and its measured duration, as well as between the former and the time it takes for $I_{\rm SXR}(t)$ to increase, from the flare onset to the moment $I_{\rm SXR}(t)$ reaches its maximum (Zhang and Dere, 2006; Mari\v{c}i\'{c} et al., 2009). Here the acceleration value was defined as the maximum ejection speed divided by the main acceleration time. It is concluded in (Temmer et al., 2008, 2010) that the time profile of the main CME acceleration a(t) is synchronised with the profile of  hard X-ray radiation intensity $I_{\rm HX}(t)$ from Reuven Ramaty High-Energy Solar Spectroscopic Imager (RHESSI; Lin et al., 2002). According to (Patsourakos et al, 2010), the profile $a(t)$ is close to the time dependence of a derivative of the soft x-ray radiation intensity $dI_{\rm SXR}/dt$. This parameter is sometimes used as an analogue of $I_{\rm HX}(t)$. This is due to the existence of the  Neupert effect (Neupert, 1968), according to which the temporal changes of $I_{\rm HX}(t)$ or of the microwave radiation intensity during the pulse phase of a flare are close to the time profile $dI_{\rm SXR}/dt$.

%{\S}{\bf --- limb CMEs} \\
CMEs for which the initial motion stage was studied and their kinematic characteristics were compared with soft and hard x-ray radiation intensity were mostly limb CMEs. Their sources are relatively close to the solar limb, and the axes of such ejections are presumably located near the sky plane. The 3-D geometrical and kinematic characteristics of such CMEs are believed to be close to those determined from sky-plane observations of these ejections, i.e. for limb CMEs, the projective effects do not make a large impact on determining their real parameter values.

  %{\S}{\bf --- halo CMEs} \\
A special group is distinguishable among all observable coronal mass ejections, called "halo" CMEs (HCMEs). They are observable in the field of view of a coronagraph as areas of enhanced brightness completely surrounding the occulting disk and expanding in all directions (Howard et al., 1982). Some HCMEs move towards the ground-based observer (frontside HCMEs), while others move in the opposite direction (backside HCMEs). In the former case, the CME sources are on the visible solar disk. It is such HCMEs that play a significant role in space weather: their influence on the Earth magnetosphere can lead to strongest geomagnetic storms (Gopalswamy, 2009).

  %{\S}{\bf --- the site of halo CME formation} \\
HCMEs are quite convenient objects for exploring the formation mechanisms for coronal mass ejections and the initial phase of their motion, especially when HCME sources are relatively not far from the center of the visible solar disk. This is due to the fact that such cases provide the opportunity to "see" all the phenomena at the site of CME formation and in the adjoining areas of the solar disk, as well as for a more correct examination of magnetic field dynamics in this area. The above merits compensate for the basic disadvantage that is inherent in studies of the  initial stage of HCME motion - increased influence of projective effects when determining the position, speed and acceleration of the leading edge (LE) of the HCME in three-dimensional space. The kinematic characteristics for the initial stage of two moving HCMEs as well as the relationship between these characteristics and the parameters of hard X-ray radiation from the HCME-related flare area have been discussed in, e.g. (Temmer et al., 2008).

  %{\S}{\bf --- CMEs moving at high speed in the field of view of the LASCO} \\
Of special interest are CMEs moving at high speed in the field of view of the Large Angle and Spectrometric Coronagraph (LASCO; Brueckner et al., 1995) onboard the Solar and Heliospheric Observatory (SOHO), and featuring a short main acceleration phase varying from a few minutes to several tens of minutes. Fast CMEs are often found to be related to to powerful flares of the M and X (X-ray) classes.

  %{\S}{\bf --- properties of six HCMEs selected from a group of fastest-moving ejections} \\
This paper inspects the properties of six HCMEs selected from a group of the fastest-moving ejections, $V>1500$ km/s, observed over solar cycle 23. The properties include the kinematics, angular sizes and trajectories of the ejections. The connection between these HCMEs and solar flares has also been examined. The initial stage of HCME motion was identified based on high temporal resolution data of the Solar X-ray Imager (SXI; Hill et al. 2005) onboard the GOES-12 space observatory. This paper relies on the SXI as an instrument with a minimum time cadence of 60 seconds and spatial resolution of 5 arcsec per pixel.

  %{\S}{\bf --- One of the reasons for selecting the fastest HCMEs} \\
One of the reasons for selecting the fastest HCMEs is that such ejections are, on the average, characterized by higher brightness than slow HCMEs both in the ejection body area and in the shock wave area. The same property is also valid for fast limb CMEs (Fainshtein, 2007). This allows for a more precise identification of the HCME body boundary and of the ejection-related shock front as well as for tracing the ejection over large distances. Moreover, the fast ejections are more often related to powerful x-ray flares and eruptive filaments. Finally, the fast HCMEs belong to more geoeffective events than do the slow HCMEs.

\section{Data and research methods} %%%%%%%%%%%%%%%%%%%%%%%%%%%%%%%%%%%%%%%%
      \label{S-general}

The properties of the following HCMEs registered in the SOHO/LASCO field of view have been studied: 29 Oct 2003 (20:54:05 UT); 15 Jan 2005 (06:30:05 UT), 15 Jan 2005 (23:06:50 UT), 17 Jan 005 (09:54:05 UT), 22 Aug 2005 (17:30:05 UT) and 23 Aug 2005 (14:54:05 UT). The time in brackets is the first registration of the ejection in the SOHO/LASCO field of view according to the catalogue \url{(http://cdaw.gsfc.nasa.gov/CME_list/HALO/halo.html)}. The events were selected according to the HCME speed value ($V>1500$ km/s) and HCME shape in the SXI field of view. The "linear" ejection speed was used as HCME speed for event selection in this case as found from a linear approximation of the positions of the fastest feature of the CME front versus time (this speed is listed in the first of three speed value columns in the above catalogue). SXI data were used to choose the HCMEs with trajectories that only relatively slightly deviated from a straight line in the SXI field of view. All the selected HCMEs are visible on the solar disk as wide loop-like structures in active regions.

  %{\S}{\bf --- Image preprocessing procedures} \\
To reliably select HCMEs in the SXI field of view, the images were subjected to some pre-processing involving the following procedures:\\
  1) Flat field correction of solar images was carried out.\\
  2) Rotating the raw solar image in order for the vertical line of this image frame to coincide with the Sun's "north - south" line.\\
  3) The solar disk centre coordinates $X_{\rm 0}$ and $Y_{\rm 0}$ were re-determined more precisely for the image. Identifying moving structures in an image series requires that all images of the series should be superimposed, which in turn demands that the solar disk center coordinates should be of high accuracy. The values of these parameters in the header of raw *.FITS files are not precise for some events, therefore $X_{\rm 0}$ and $Y_{\rm 0}$ were re-determined with more precision.\\
  4) Superimposing the nearest-in-time images. A high-brightness area adjacent to the active region under inspection but featuring no flares, ejections or appreciable movements during the relevant observation period were identified i n each solar disk image. Analysis of the 2D-crosscorrelation function of the selected elements of two nearest-in-time images enabled their displacement $\Delta x_{\rm 0}$ and $\Delta y_{\rm 0}$ to be found. The image corresponding to the second moment of time was displaced based on the $\Delta x_{\rm 0}$ and $\Delta y_{\rm 0}$ values. The third image of the series was displaced $(\Delta x_{\rm 0}+\Delta x_{\rm 1})$ and $(\Delta y_{\rm 0}+\Delta y_{\rm 1})$, where $\Delta x_{\rm 1}$, $\Delta y_{\rm 1}$ are the horizontal and vertical displacement found for overlapping the third image relative to the second image of the series and so on.\\
  5) Running-average filtering of data in order to decrease high-frequency noise.\\
  6) Calibrating the active region images for all image series for each event under analysis. The procedure consisted in the following: the average intensity value $I_{\rm 0},I_{\rm 1},...I_{\rm n}$ were found for the nearest quiet regions, for each active region image. The intensity value in each pixel of the series was multiplied by a corresponding factor $k_{\rm 1}=I_{\rm 0}/I_{\rm 1},...,k_{\rm n}=I_{\rm 0}/I_{\rm n}$,  where $0,1,...n$ is the series image index.\\
  7) Obtaining brightness ratios for the two nearest-in-time images of the active region in question. THe active region image for each moment of time was divided by the image for the previous moment of time. We will hereafter use the term of 'running' ratio images between nearest-in-time frames for this procedure.

  %{\S}{\bf --- The HCME front top, middle and basis were identified in these brightness distributions} \\
To find the coordinates, speed and acceleration versus time of the leading edge (LE) of the HCMR, brightness distributions were constructed along the HCME axis (see Figure 1 (A,B) in subsection 3.1, showing pre-processing fragments of solar disk images), Figure 1(D,E). These distributions refer to the straight line drawn from the flare centre through the fastest area in the leading edge of the HCME. The HCME front top, middle and basis were identified in these brightness distributions (Figure 1(B)). Time dependences of their speed $V_{\rm i_-(i+1)}(t)$ were plotted for each of these singularities of the leading edge, where $i$ and $(i+1)$ are the numbers of the two nearest-in-time images  at time moments $t_{\rm i}$ and $t_{\rm (i+1)}$. When combining SXI data with LASCO data we only used speeds obtained in the front of the base of the HCME.

  %{\S}{\bf --- small deviations of the ejection trajectory} \\
Sky-plane distances $dL_{\rm i_-(i+1)}$ between the coordinates of the HCME front basis at consecutive moments of time ti and ti+1 were determined in order to find the HCME front speed $V_{\rm i_-(i+1)}(t)$. In contrast to many other papers, we did not ignore small deviations of the ejection trajectory from a straight line at the initial stage of motion, therefore in our case $dL_{\rm i_-(i+1)}$ is the length of the curvilinear trajectory along which the HCME front travels over time interval ($t_{\rm i}$, $t_{\rm (i+1)}$). The average speed in this part of the trajectory was found from the relationship:

\begin{equation}\label{Eq-1}
V_{\rm i_-(i+1)}\equiv V_{\rm (t_i+t_{i+1})/2}=dL_{\rm i_-(i+1)}/(t_{\rm (i+1)}-t_{\rm i})
\end{equation}

  %{\S}{\bf --- the speed values found using two nearest nearest-in-time images} \\
It follows from (1) that the speed value $V_{\rm i_-(i+1)}$ found using the two nearest nearest-in-time images of the solar disk at moments $t_{\rm i}$ and $t_{\rm i+1}$ is attributed to moment $(t_{\rm i}+t_{\rm (i+1)})/2$.

  %{\S}{\bf --- intermediate speeds} \\
While plotting $V_{\rm i_-(i+1)}(t)$ we also calculated intermediate speeds based on the last coordinates of the HCME front as found in SXI data and the first coordinates of the front based on LASCO data taken from the catalogue at \url{(http://cdaw.gsfc.nasa.gov/CME_list/HALO/halo.html)}. For the 29.10.2003 (20:54:05 UT) event, points obtained from the MarkIV K-coronameter of the Mauna Loa Solar Observatory (MLSO), \url{(http://mlso.hao.ucar.edu/)}, were added onto the speed profile $V_{\rm i_-(i+1)}(t)$.

  %{\S}{\bf --- the minimum error in speed determination} \\
The set of points $V_{\rm (t_i+t_{(i+1)})/2}$, with the measurment error taken into account, was used for finding continuous dependence $V(t)$ by means of a B-spline interpolation of these values. The minimum error in speed determination was found from this relationship (Shanmugaraju et al., 2010):

\begin{equation}\label{Eq-2}
\delta V = 2\delta L/(t_{\rm (i+1)}-t_{\rm i})
\end{equation}
\\
  %{\S}{\bf --- real velocity dependence was effectively set as a discrete set of speed values} \\
where $\delta L$ is the spatial resolution of the telescope. The real error of determining speed $V_{\rm i_-(i+1)}$ was, as a rule, more than $2\delta L/(t_{\rm (i+1)}-t_{\rm i})$. In that case the error of determining speed $V_{\rm i_-(i+1)}$ based on SXI data was found using statistical methods of analysing the value of $V^k_{\rm i_-(i+1)}$, $k=1-N$ obtained from repeated measurements ($N$ is the total number of velocity measurements) of the HCME front coordinates in each solar disk image. For some events in the field of view of the LASCO C2 coronagraph, the error $\Delta L$ of determining the HCME front coordinates in the LASCO C2 field of view was set at $\pm 5$ pixels at $R<5R_{\rm o}$, by analogy with (Gallagher et al., 2003), in order to obtain a monotonous dependence $V(t)$ passing through all values $V_{\rm i_-(i+1)}$ with the error taken into account. For the LASCO data, $\Delta L\sim 0.5R_{\rm o}$ at $R>5R_{\rm o}$, where $R_{\rm o}$ is the solar radius, (Shanmugaraju et al., 2010). The dependence $V(t)$ was effectively set as a discrete set of speed values with a small time step that was much smaller than $(t_{\rm (i+1)}-t_{\rm i})$.

  %{\S}{\bf --- acceleration values} \\
Acceleration $a(t)$ was found from the relationship:

\begin{equation}\label{Eq-3}
a(t) = dV(t)/dt
\end{equation}

  %{\S}{\bf --- the precision of finding of velocity and acceleration values} \\
Note that the precision of finding $V(t)$ and $a(t)$ also depends on the accuracy of identifying the same element in coronal images for various moments of time (see Vr\v{s}nak et al, 2007).

  %{\S}{\bf --- the motion of an element on the HCME axis} \\
In this paper we analyzed the motion of an element on the HCME axis, i.e. a straight line from the probable centre of the HCME source through the middle between the extreme-in-latitude ejection points. Other ways are also possible for identifying the HCME element to examine its motion (for example the farthest HCME area from the ejection origin in each image).

  %{\S}{\bf --- telescopes for identifying loop-like structures} \\
SOHO-based Extreme-Ultraviolet Imaging Telescope (EIT; Delaboudiniere et al., 1995) data were used for identifying loop-like structures that could form the flux-rope basis of a would-be HCME. Data from ground-based telescopes observing the Sun in $H\alpha$ \url{(http://swrl.njit.edu/ghn_web/)}, the PICS instrument (MLSO), \url{(http://mlso.hao.ucar.edu/mlso_data_PICS_2005.html)} and Culgoora\linebreak observatory, \url{(ftp://ftp.ips.gov.au/wdc-data/solar/data/culgoora/)} were used to identify eruptive filaments (prominences) in the HCME origin which could be related to HCME emergence.

  %{\S}{\bf --- the kinematics of the HCME-related shock and of the HCME body} \\
This paper compares the kinematics of the HCME-related shock and of the HCME body for the 15.01.2005 (23:06:50 UT) event in the LASCO C2 and C3 coronagraph field of view. Two approaches were employed for shock wave identification. In the first case, radial brightness distributions of running difference images between subsequent images were obtained by subtracting the image at the earlier moment of time $t_{\rm 0}$ from the image at point of time $t_{\rm 1}$. If a brightness jump on the difference brightness images forms  at the HCME boundary, with a spatial scale of approximately $(1-2.5)\delta L$, where $\delta L$ is the spatial resolution of the coronagraph ($\delta L=0.025R_{\rm o}$ for LASCO C2 and $\delta L=0.125R_{\rm o}$ for LASCO C3), not exceeding a certain limit value, then this brightness jump was accepted as the shock wave front. When due to their small amplitude, it was not possible to identify sharp brightness jumps against the noise background, then the rather sharp border of the rarefied area visually observable before the HCME body on running subtraction images was accepted as shock wave front. Examples illustrating the two methods of shock wave identification are given in section 3. It is also shown there how the HCME body was identified.

\section{Results of Analysis of the kinematics properties of halo coronal mass ejection} %%%%%%%%%%%%%%%%%%%%%%%%%%%%%%%%%%%%%%%%
      \label{S-features}

\subsection{17 Jan 2005 HCME event} %%%%%%%%%%%%%%
  \label{S-equations}
This event has already been analyzed in (Vr\v{s}nag et al., 2007; Temmer et al, 2008). Given the above-discussed features of image preprocessing and LE ejection position determination at various moments of time, the results we have obtained have supplemented the results in the above-cited papers. Additionally, we have discussed earlier-unaddressed properties of this ejection.

  %{\S}{\bf --- black and white loop-like structure} \\
This HCME was first registered in the LASCO C2 field-of-view 17.01.2005 (09:54:05 UT). The "Linear" speed of ejection was $2547~km/s$ according to the catalogue \url{(http://cdaw.gsfc.nasa.gov/CME_list/HALO/halo.html)}. GOES-12 registered a rapid (within approximately 10 minutes) increase in SXR intensity starting from the flare onset to the moment ISXR achieved its maximum value. According to RHESSI, this flare was accompanied by powerful streams of hard X-ray radiation (HXR) including a photon energy range of about $50-100~keV$.

  %{\S}{\bf --- first registered in the LASCO С2} \\
Figure 1(A,B) shows examples of solar area images from SXI data where a moving light-coloured loop-like structure is clearly observed with its LE designated as $Fw_{\rm 1}$. Similar to the authors of (Temmer et al, 2008) we believe this structure to be a coronal mass ejection, the front of which is marked with an arrow labelled $Fw_{\rm 1}$ in the LASCO C2 field of view (Figure 1(C)). Figure 1(A,B) also shows straight lines from the spot marked with a dagger which approximately corresponds to the ejection origin, as well as those passing through the middles of LE's of "white" loop-like structures. Figure 1(D,E) displays brightness distributions along these straight lines in the areas of the dark and light structures of the ejection, for various moments of time. The ejection front top, middle and basis are marked (top to bottom) with daggers. As has already been noted in section 2, the basis of its front was used for plotting the kinematic characteristics of the motion of "white" loop-like structures. The dark area below "white" loop-like structures corresponds to the position of a "white" loop at a previous moment of time. We also used the forward boundary of the dark area for plotting the kinematic characteristics of the HCME during the initial stage of its motion.

  %{\S}{\bf --- another moving loop-like structure} \\
Apart from the moving structure with its leading edge labelled $Fw_{\rm 1}$, another moving loop-like structure can be observed in Figure 1(A,B), its LE  labelled $Fw_{\rm 2}$. This structure may be linked to another HCME, labelled $Fw_{\rm 2}$ in Figure 1(C). Note, however, that no coronal mass ejection corresponding to this loop-like structure is to be found in the catalogue \url{(http://cdaw.gsfc.nasa.gov/CME_list/)}. This may be due to the fact that what we are dealing with are not two different CMEs, but a compound HCME with the loop-like structures with their fronts labelled $Fw_{\rm 1}$ and $Fw_{\rm 2}$ being its separate parts.

\begin{figure}    %%%%%%%%%%%%%%%%%% FIGURE 1
   \centerline{\includegraphics[width=1.0\textwidth,clip=]{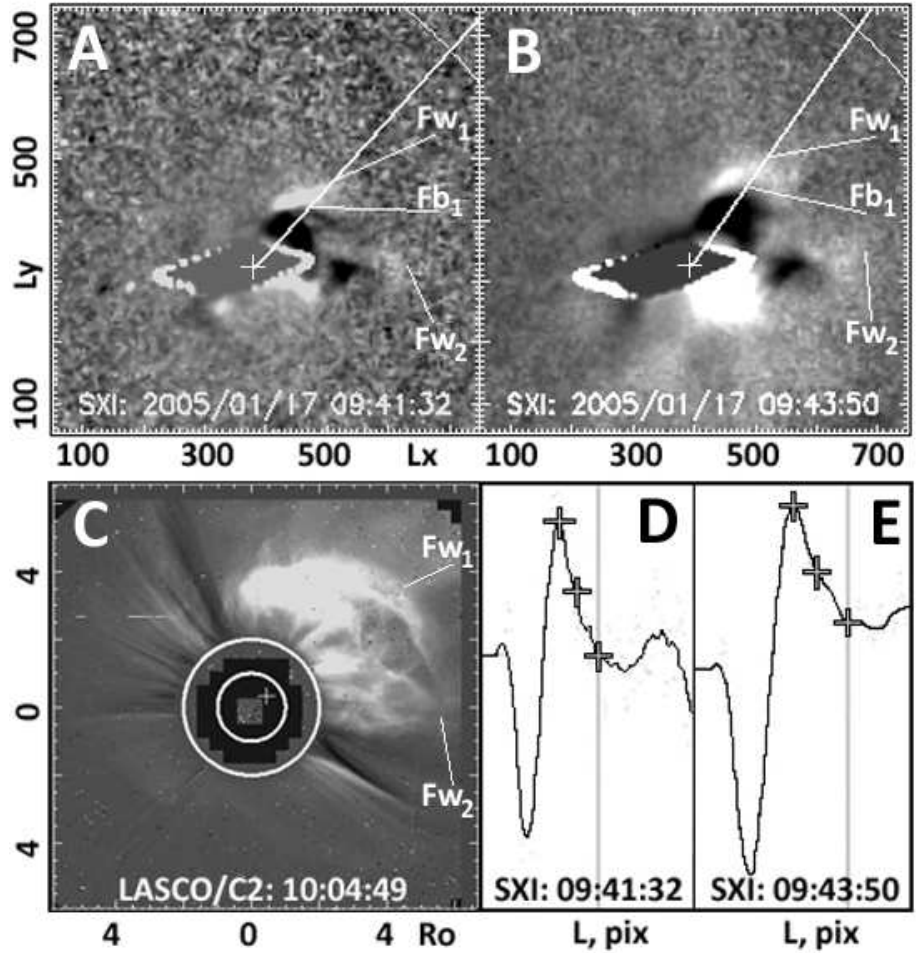}}
   \caption{(A), (B) present the images of solar disk fragments from the SXI data obtained by moving division of adjacent images (see explanations in section 2). The dagger marks the CME source proposed location. $Fw_{\rm 1}$ and $Fw_{\rm 2}$ mean the leading edge (LE) of white loop-like structures, $Fb_{\rm 1}$ is the leading edge of the dark structure accompanying the white loop-like structures. Straight lines from the ejection origin location passing through its LE middle are the brightness scan lines. Distances from the solar disk center $L_{\rm x}$, $L_{\rm y}$ in seconds of arc are shown on the axes; (C) presents the CME images within the LASCO C2 field-of-view. Here, the distances are expressed in the solar disk radii, Ro. White circles highlight the solar disk boundaries and the C2 artificial moon; (D), (E) present the brightness distributions along the straight lines from the CME proposed location in Figures (A) and (B). The daggers in these distributions mark (top to bottom) the LE apex, middle and foot-point of the white loop-like structure identified with the CME. The distances expressed in the SXI radiation recorder pixels are marked on the abscissa axis. We did not cite the brightness values in these Figures.}
   \label{F-simple}
\end{figure}

  %{\S}{\bf --- the time velocity profile} \\
The solid curve in this figure is the time profile $V(t)$ of the ejection speed based on combined SXI and LASCO data. Additionally, the dashed line in Figure 2(A) shows, for comparison, the time dependence of soft X-ray radiation intensity $I_{\rm SXR}(t)$ from the HCME-related flare area based on GOES-12 data.

\begin{figure}    %%%%%%%%%%%%%%%%%% FIGURE 2
   \centerline{\includegraphics[width=1.0\textwidth,clip=]{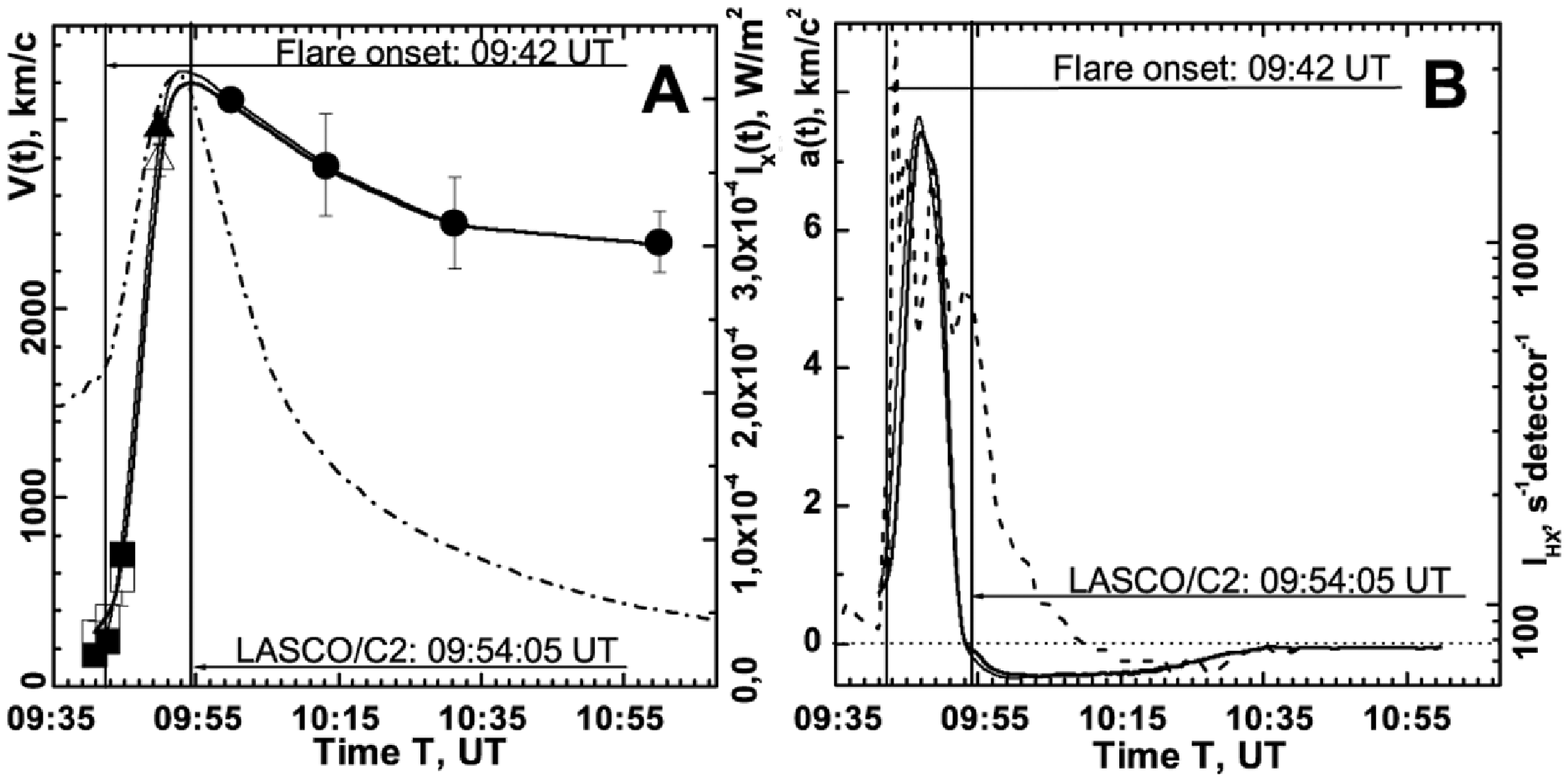}}
   \caption{(A) presents the 23 Aug 2005 HCME velocity  compared with the soft X-ray intensity ISXR(t). White squares show the velocity values determined by the brightness profile base in the "white" loop-like structures' region, black squares show the velocity values determined by the "black" loop-like structures' apices, black circles show the velocity values from the LASCO data, white and black triangles mark the velocity intermediate values obtained by the SXI last image and by the C2 first image for "white" and "black" loop-like structures, respectively; the dash-dotted line shows the soft X-ray intensity profile $I_{\rm SXR}(t)$ obtained from GOES-12. (B) presents the HCME a(t) acceleration compared with the hard X-ray intensity $I_{\rm HX}(t)$. Solid thin and thick lines show the CME acceleration profiles corresponding separately to the "white" and "black" loop-like structures observed by SXI; the dashed line shows the 50-100 keV hard X-ray intensity values $I_{\rm HR}$ from RHESSI. Figures (A) and (B) present the instants corresponding to the flare onset and the CME first observation within the LASCO C2 field-of-view.}
   \label{F-simple}
\end{figure}

  %{\S}{\bf --- another moving loop-like structure} \\
In a Figure 2(B) the time profile of HCME acceleration $a(t)$ and time dependence of intensity of hard X-ray radiation $I_{\rm HXR}(t)$ in energy range of $50-100~keV$ for comparison is shown.

  %{\S}{\bf --- conclusions from the graphic} \\
It follows from Figure 2 that:\\
  1) HCME starts to move along minutes prior to the onset of the relevant flare.\\
  2) HCME speed quickly reaches a maximum value and then decreases by $\approx 800~km/s$ within $\approx 40$ minutes, in the LASCO field of view, further continuing to decrease with time but at a significantly slower rate.\\
  3) The temporal variation of $V(t)$ is synchronized with the $I_{\rm SXR}(t)$ variation. $V(t)$ reaches its maximum $\sim 1$ minute after maximum $I_{\rm SXR}(t)$. This characteristic agreement between $V(t)$ and $I_{\rm SXR}(t)$ as well as other HCME characteristics - and those pertaining to radiation from the flare area - are listed in Table 1 for all the events in question.\\
  4) The main acceleration of HCMEs is characterized by a bell-shaped curve with a rapid increas and rapid decrease. The acceleration becomes negative in the LASCO C2 and C3 fields of view, its absolute value decreasing with time. Such behavior agrees qualitatively with the concept of the ejection expanding in the self-similar mode during this period (Uralov, Grechnev, Hudson, 2006). The positive-to-negative acceleration reversal occurs approximately in the middle of the time interval during which intensity $I_{\rm HXR}$ remains at a conditionally chosen level $I_{\rm HXR}=100$ hard X-ray intensity measurement units.

  %{\S}{\bf --- maximum HCME speed} \\
Let us pay attention to the fact that maximum HCME speed $V_{\rm MAX}$ was $3200~km/s$. Given this is the speed of an ejection as projected onto the skyplane, the HCME speed along the axis may be assumed to exceed $4570~km/s$ in 3-D space which is consistent with the CME axis deviating no more than $45^\circ$ from the radial direction.

\begin{figure}    %%%%%%%%%%%%%%%%%% FIGURE 3
   \centerline{\includegraphics[width=1.0\textwidth,clip=]{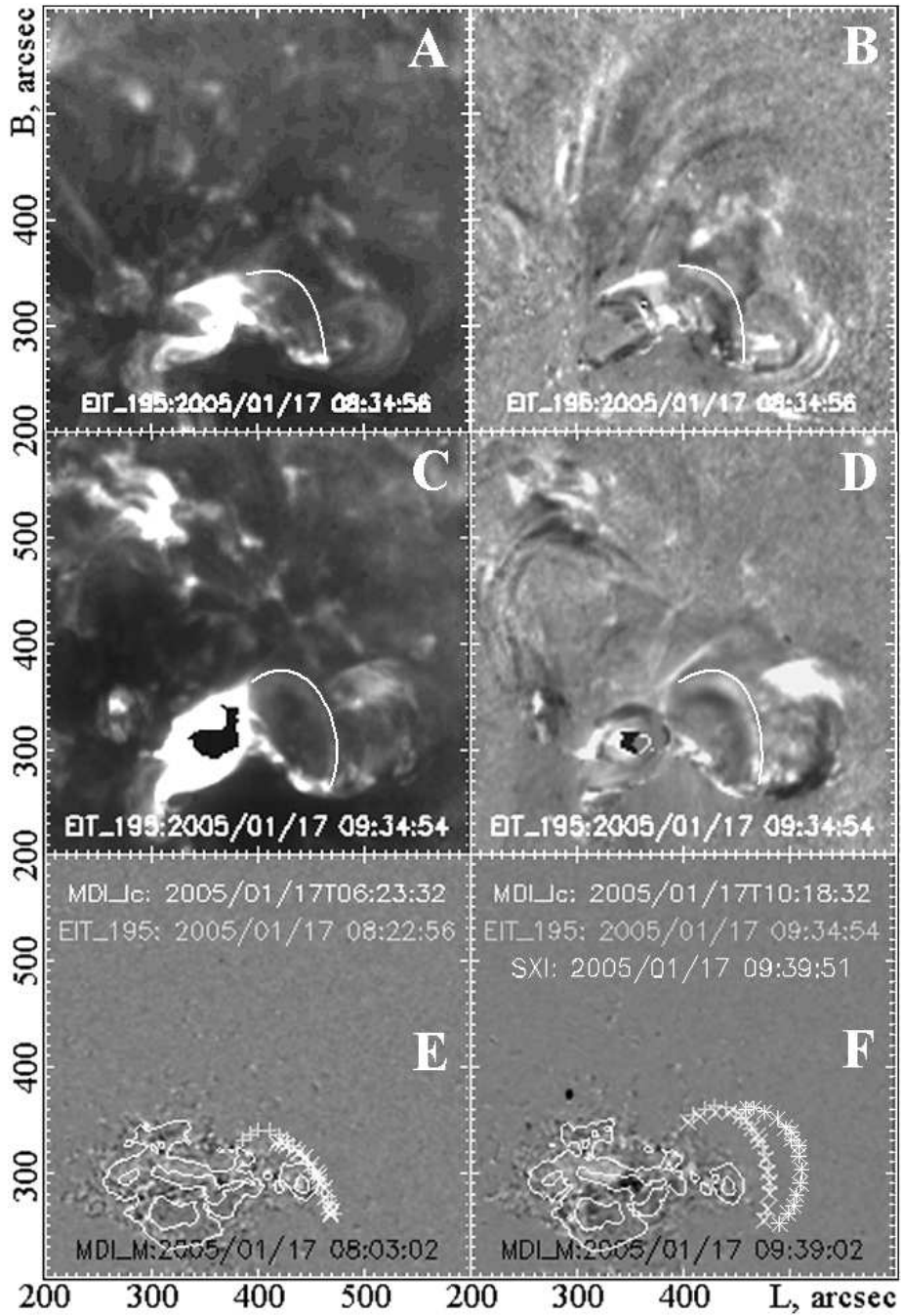}}
   \caption{(A), (B) - Illustration of the 19.5 nm Fe XII line loop-like structure existence (its boundary is denoted by a white line) observed within the SXI field-of-sight long before the HCME beginning and that will have generated with time the HCME loop. (C) – the raw EIT image NOAA 10720, (D) – the running ratio image (E), (F) - – the running ratio magnetogram; the coordinates of loop-like structures in different wavelengths at different points of time were plotted over the image. Pluses are the coordinates of the loop-like structure in raw images on EIT image at $\lambda=19.5~nm$. Crosses are the coordinates of the same loop-like structure on running ratio EIT images. Asterixs show the loop-like structure boundary on SXI images.}
   \label{F-simple}
\end{figure}

  %{\S}{\bf --- loop-like structure on EIT images} \\
It is impossible to derive from SXI data how exactly the loop-like structure which is in the HCME basis formed. Some information on this has been successfully obtained from solar images in the Fe XII line at $\lambda =19.5$ nanometers based on SOHO/EIT data. It turns out that the loop-like structure which was to form the future HCME existed long before the moment when it was registered in the SXI, Figure 3(A-C). Analysis of $H\alpha$ observations in observatories \url{(http://swrl.njit.edu/ghn_web/)}, \url{(http://www.ips.gov.au/Solar/2/1)} and by the PICS telescope (MLSO) has shown that this loop-like structure is not a filament and hence consists of hot coronal plasma. This loop-like structure has begun a slow forward motion long before the moment of its first registration in the SXI field of view. Average speeds of this structure over the period between $t_{\rm (SXI-1)}=...$ and the moment  nearest to it $t_{\rm EIT}...$ when the loop was registered according to EIT were $<V_{\rm EIT}>= 4.4~km/s$.

  %{\S}{\bf --- selected loop-like structure on EIT images} \\
Since the set of emission loops are simultaneously registered in the active region on Sun's images in the extreme ultraviolet line, then we used the following referred to below for finding the loop-like structure (the beginning of the HCME). First of all, there were heliographic coordinates of the leading edge points of HCME in field of view of SXI. The coordinates of the leading edge of the loop-like structure were plotted on the Sun image in the ultraviolet line at  $\lambda =19.5~nm$ (EIT data) at the moment of time, as close as possible to the first moment of HCME observation in the field of view of SXI. After this, the emission loop closest in position, in form and in arrangement to the loop-legs of the LE of the ejection was found on the EIT image. It appears that the loops marked out in such a way on EIT images look less bright, than other emission loops in the active region investigated. In some cases (see the following events) one part of the loop-like structure was well observed in the extreme ultraviolet line, the other in a soft x-ray range. And, finally, we note that all the selected loop-like structures in the extreme ultraviolet range were wider than the adjacent emission loops in the active region.

\subsection{23 Aug 2005 HCME event} %%%%%%%%%%%%%%
  \label{S-equations}
Figures 4 and 5 displays images of part of the solar disk and plots for the 23 Aug 2005 HCME similar to the images and plots for the ejection registered on 17 Jan 2005. This ejection has emerged in NOAA 10798 and was related to an X-ray class M2.7 flare with heliographic coordinates S14W90, i.e. located close to the limb. The halo (i.e. areas of high brightness round the occulting disk of a coronagraph) formation mechanism in this event most likely differs from the halo formation mechanism for the ejection observed 17 Jan 2005 (Gopalswamy et al., 2010). In the former case the HCME-related flare and the origin of the ejection were near the solar disk centre (Figure 1,3). Therefore it is feasible to believe that the halo for the 17.01.2005 event was formed by luminescence from the HCME body and shock-compressed plasma behind the front of the ejection-related shock wave. In the latter case the halo is formed by the great shock-compressed area of plasma behind the shock wave front and/or by an EIT wave from the HCME formation area (the latter possibility was suggested by V.V. Grechnev in private communication).

\begin{figure}    %%%%%%%%%%%%%%%%%% FIGURE 4
   \centerline{\includegraphics[width=1.0\textwidth,clip=]{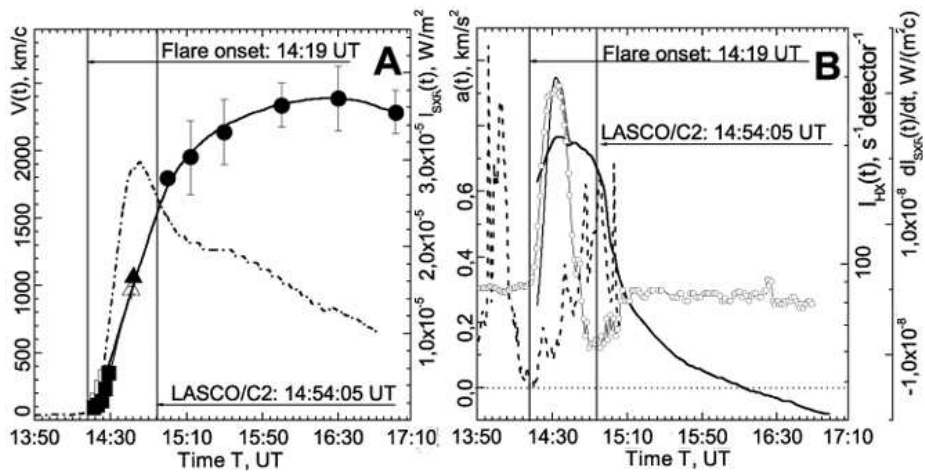}}
   \caption{The same as in Figure 2, but for the 23 Aug 2005 HCME event. The circles in Figure (B) additionally show the time dependence $dI_{\rm SXR}(t)/dt$.}
   \label{F-simple}
\end{figure}

  %{\S}{\bf --- Calibration of the observed brightness} \\
The 23 Aug 2005 ejection motion kinematic properties turned out to differ substantially from the corresponding properties of the 17 Jan 2005 HCME. In this case, after a short period of rapid increase for about 50 minutes, the HCME velocity continued to increase, though rather slowly, and upon reaching its maximal value, it started to diminish. Here, we conditionally accepted $\approx 15:10$ UT as the HCME main acceleration domain end. This ejection acceleration amplitude turned out to be an order of magnitude less than that for the 17 Jan 2005 HCME. The 23 Aug 2005 HCME acceleration profile also differs significantly from the 17 Jan 2005 $a(t)$ dependence. Upon reaching the maximal value, the HCME acceleration diminishes within $\approx 102$ minutes to zero value, and then, within negative values, it slowly increases modulo.

  %{\S}{\bf --- the motion kinematic characteristics of 23.08.2005 HCME event} \\
The association of the motion kinematic characteristics with the radiation intensity in the soft and hard X-ray bands from the associated flare domain (see Figure 4) also turned out quite different for the given HCME compared with the 17 Jan 2005 ejection. The synchronism of the ejection’s motion and the time variations for the X-ray intensities turned out to be dramatically violated. During the ejection main acceleration, the growth velocity of $V(t)$ dependence was noticeably less than the increased velocity $I_{\rm SXR}(t)$, the $V(t)$ maximum being 104 minutes after the soft X-ray radiation intensity had become maximal. The acceleration maximum $a_{\rm MAX}$ turned out close in time to reaching the maximal value $dI_{\rm SXR}(t)/dt$ (a similar situation occurred in the previous event also), and, simultaneously, it was located between the hard X-ray intensity peaks $I_{\rm HXR}(t)$ which is relatively close to the $I_{\rm HXR}(t)$ minimum. Here, we might be dealing with the case of when the Neupert effect (1968) does not work. We will discuss possible causes for the difference between the kinematics of the two HCMEs under consideration later.

\begin{figure}    %%%%%%%%%%%%%%%%%% FIGURE 5
   \centerline{\includegraphics[width=1.0\textwidth,clip=]{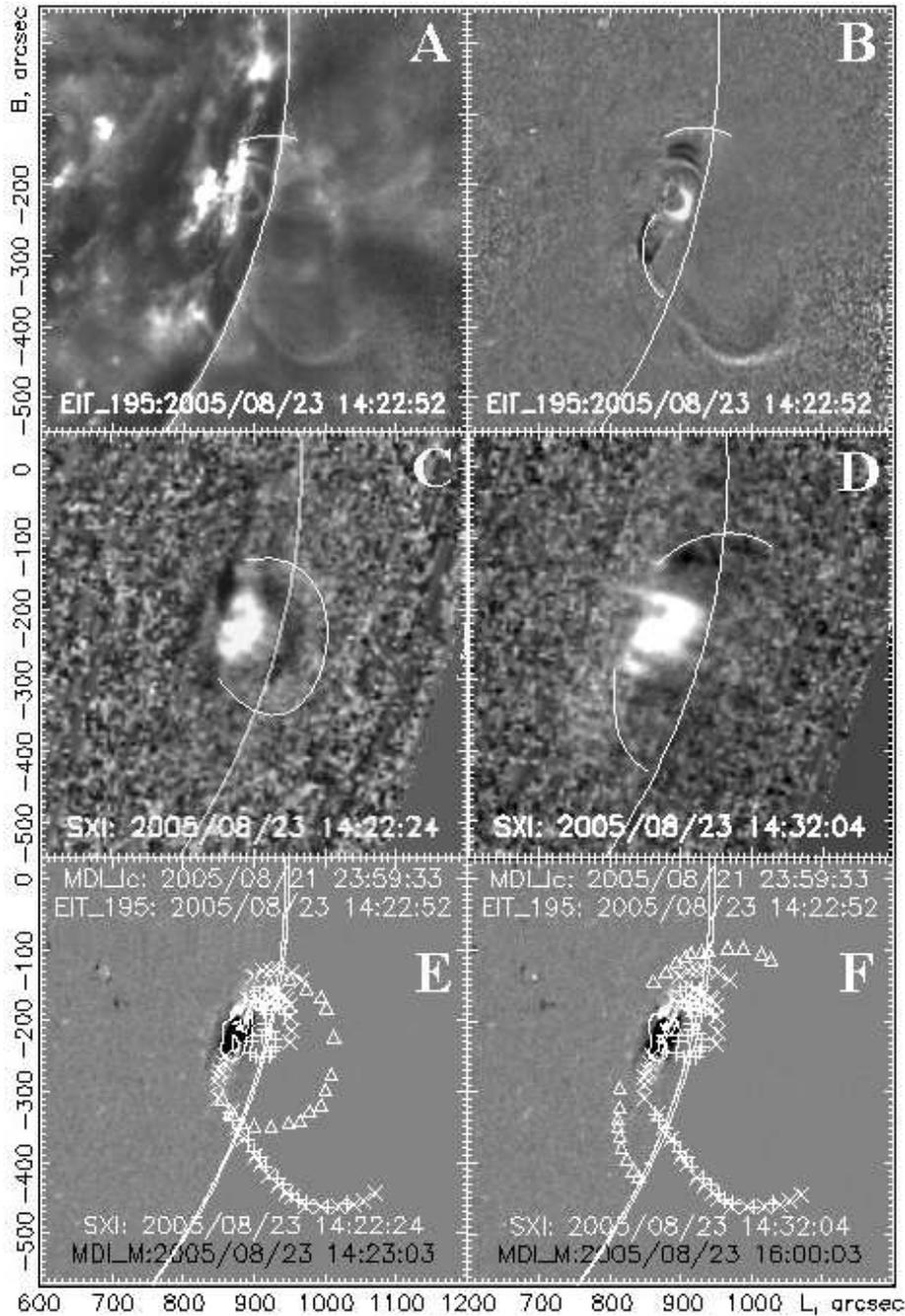}}
   \caption{The same as in Figure 3, but for the 23 Aug 2005 event. (A) – the raw EIR image of active region NOAA 10486, (B) – the running ratio EIT image, (C), (D) – the running ratio SXI images a different point in time, (E), (F) – the loop-like structure boundaries in the field-of-view of EIT (pluses and asterics) and SXI (triangles) telescopes plotted on the MDI magnetogram of this active region. Isolines show the boundary of umbra and penumbra defined according to the MDI at continuum.}
   \label{F-simple}
\end{figure}

  %{\S}{\bf --- the start of the translational motion of the 23 Aug 2005 HCME} \\
From Figure 5(A) it also follows that the translational motion of the given HCME started almost simultaneously with the flare onset. However, the loop-like structure that we associate with the 23 Aug 2005 CME was observed prior to the flare onset in active region NOAA 10798 at 13:58:59 UT from EIT. From Figure 5(A) it follows that the loop-like structure recorded in the $\lambda =19.5~nm$ huv line corresponds to the loop-like structure (the beginning of the HCME) observed in the SXI images earlier ($\sim 10$ minutes prior to the flare).

\subsection{29 Oct 2003 Event} %%%%%%%%%%%%%%
  \label{S-equations}
Figures 6 and 7 illustrate this event. The coronal mass ejection originated in the active region NOAA 10486. The ejection was accompanied by a powerful X-ray and X10/2B optical flares with the S15W02 coordinates. A peculiarity of this event is that we were offered an opportunity to see one of the HCME formation stages from SXI. The ejection formation starts with the translational motion of three loop-like structures closely spaced to one another (see Figure 6(A)). This is possibly a projection of a loop arcade rather than of isolated loops onto the solar disk. After a while following the translational motion onset, these loops form a wide front ejection integrated structure whose motion we consider as the formed CME motion. In the LASCO C2 field-of-view, the CME under study was recorded, for the first time, at 20:54:05 UT and resembled a limb CME. The CME motion direction coincided with the motion direction of the loop-like structure seen in the SXI images. And it is only within the LASCO C3 field-of-view that the CME became a halo coronal mass ejection.

\begin{figure}    %%%%%%%%%%%%%%%%%% FIGURE 6
   \centerline{\includegraphics[width=1.0\textwidth,clip=]{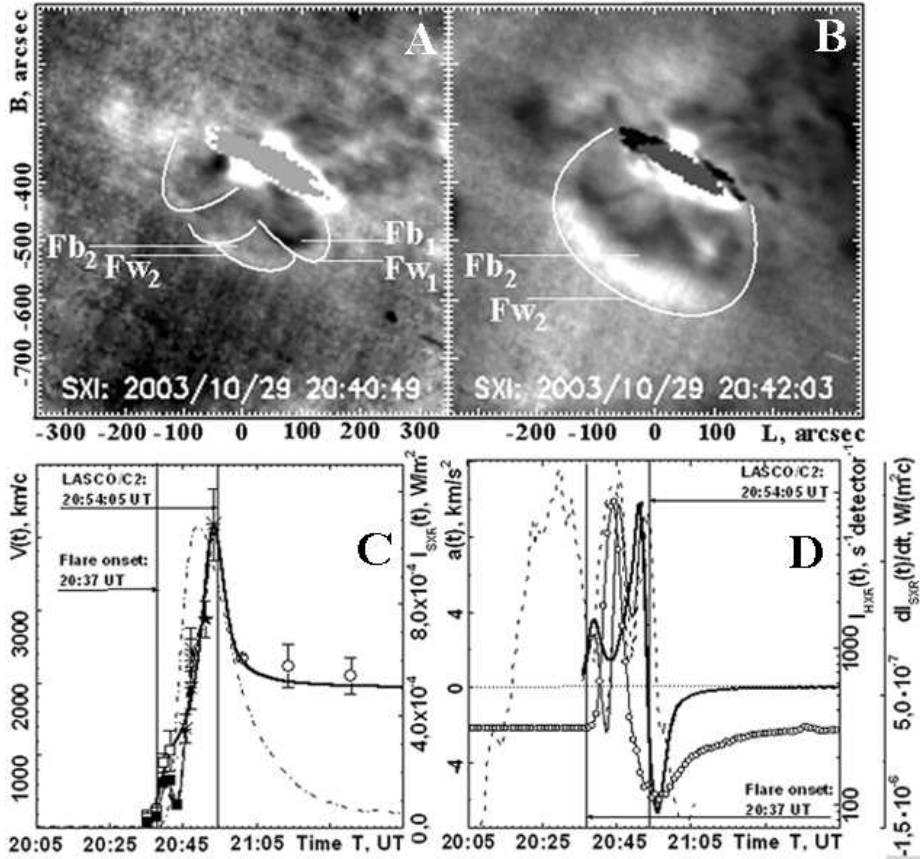}}
   \caption{(A), (B) – image of arcade of loop-like structures for the 29 Oct 2003 event in the field-of-view of SXI telescope  at a different point in time, (C), (D) - the same as in Figure 2, but for the 29 Oct 2003 event.}
   \label{F-simple}
\end{figure}

  %{\S}{\bf --- the velocity time profile} \\
The 29 Oct 2003 HCME velocity $V(t)$ time profile is similar to the corresponding velocity profile for the 17 Jan 2005 ejection: the velocity reaches its maximal value of d $4200~km/s$ rapidly (for about 38 minutes), then rapidly (for about 22 minutes) diminishes to d $2000~km/s$, and, then continues to slowly diminish (Figure 6(C)).  The acceleration profile looks rather complicated: with two peaks, and also with a slight drop at the ejection motion start and with a drop after the HCME main acceleration end, Figure 6(D). Then, as in  the 17 Jan 2005 HCME case, the ejection acceleration becomes negative and later diminishes modulo.

  %{\S}{\bf --- HCME kinematics and X-ray intesity} \\
This HCME kinematics is synchronized relatively well with the time variation for the X-ray from the associated flare domain. Velocity $V(t)$ and the soft X-ray intensity $I_{\rm SXR}(t)$ increase fast, and the ejection velocity reaches its maximum $\sim 10$ minutes after reaching the $I_{\rm SXR}(t)$ maximum. The acceleration highest peak maximum is reached d 3 minutes after the soft X-ray intensity derivative $dI_{\rm SXR}(t)/dt$ assumes maximal value, and $\approx 5$ minutes after the hard X-ray intensity $I_{\rm HXR}(t)$ middle peak maximum. As seen from Figure 6(D), the $I_{\rm HXR}(t)$ profile does not repeat the $dI_{\rm SXR}(t)/dt$ dependence consisting of one peak, but represents a 3-peak structure. This HCME translational motion starts d 3-4 minutes earlier than the associated flare onset in active region NOAA 10486.

\begin{figure}    %%%%%%%%%%%%%%%%%% FIGURE 7
   \centerline{\includegraphics[width=0.9\textwidth,clip=]{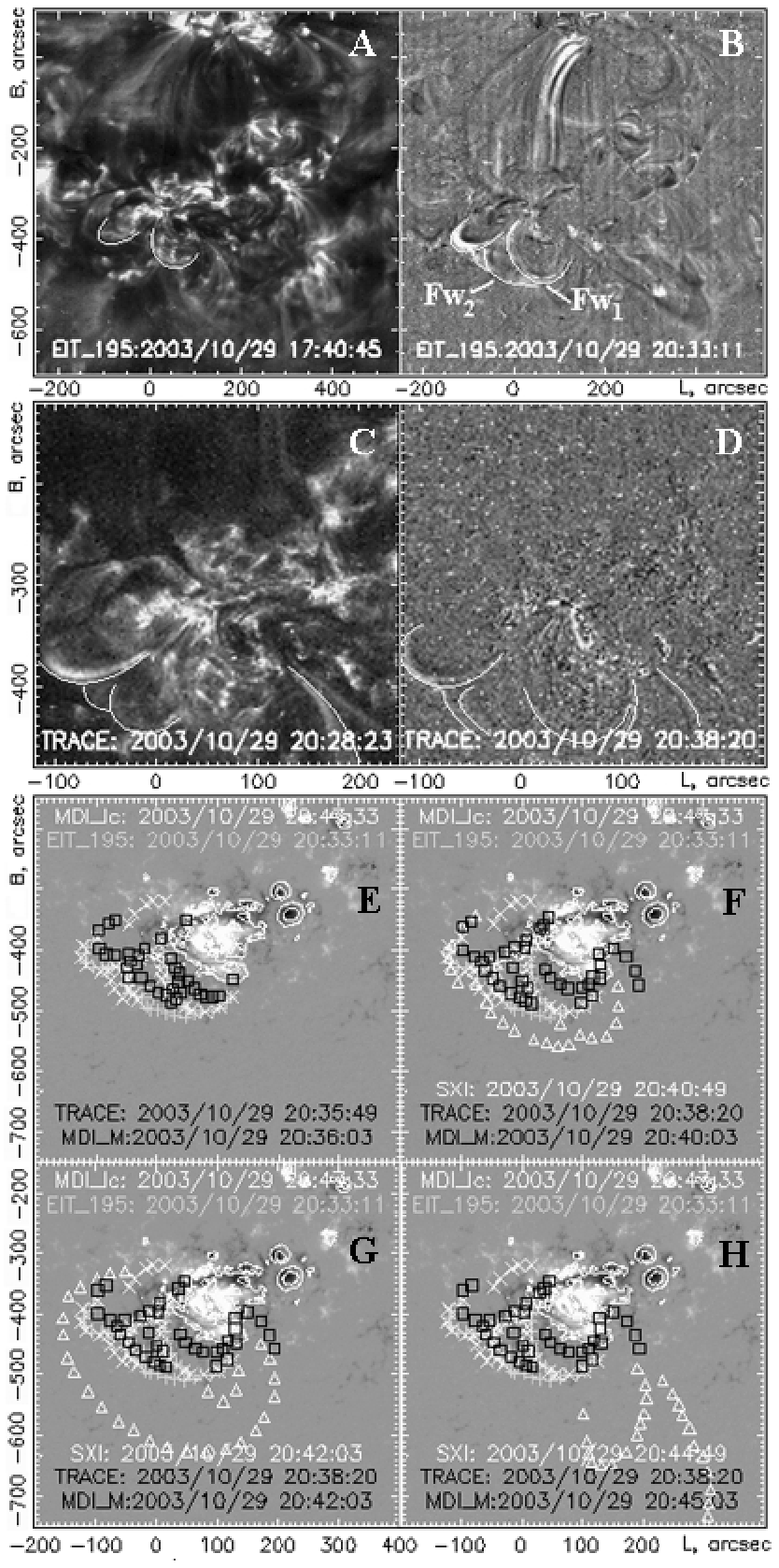}}
   \caption{((A-C) - the same as in Figure 3 (E), (F), but for the 29 Oct 2003 event using the TRACE images. The boundaries of loop-like structures forming the HCME are marked by white lines. (D-F) - the same as in Figure 3(A), (B), but for the 29 Oct 2003 event. (G-I) are similar to Figure 5(E), (F); the boundaries of the loop arcade or loop-like structures are marked by squares on TRACE images.}
   \label{F-simple}
\end{figure}

  %{\S}{\bf --- HCME on TRACE images} \\
As in the two previous events under consideration, in the course of the 29 Oct 2003 event, one observed a few traveling loop-like structures, the predecessors of the loops that were observed with SXI (Figure 7), within the EIT and Transition Region and Coronal Explorer (TRACE; Handy et al., 1999) fields-of-view in the huv emission lines. However, we managed to record the formation of one wide-front loop-like structure enveloping several loops (a loop arcade) only from the SXI data.

\subsection{On the existence of two CME types depending on the velocity V(t) profile} %%%%%%%%%%%%%%
  \label{S-equations}
Figure 8 presents the $V(t)$ profiles for all the HCMEs under consideration. One can see that the velocity of four ejections reached its maximum rapidly, and then in a few tens of minutes it diminished by several hundreds of km/s, and, later, continued to diminish, though rather slowly (Figures 8(A)). For two ejections, the time-dependant velocity variation was totally different (Figures 8(B)). The ejection velocity augmented rapidly at its motion start, and then continued to augment, but much more slowly. In this case, the HCME velocity reached its maximum within the LASCO field-of-view circa two hours after the ejection motion start.

\begin{figure}    %%%%%%%%%%%%%%%%%% FIGURE 8
   \centerline{\includegraphics[width=1.0\textwidth,clip=]{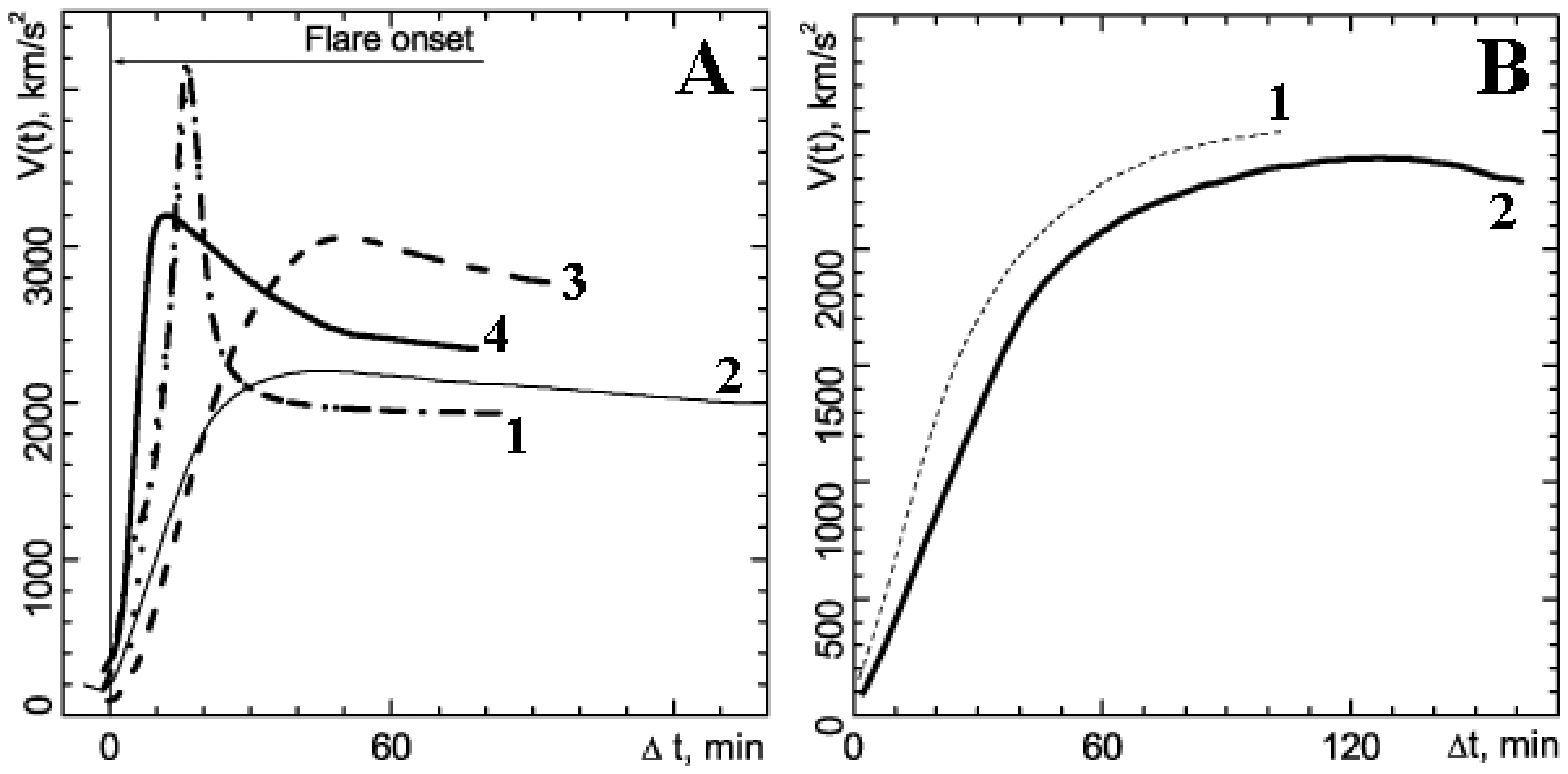}}
   \caption{Illustration of the two types of HCMEs differing in time profile V(t). (A) - for I HCME kinematics type (where 1 - for 29 Oct 2003 event, 2 - for 15 Jan 2005 event (05:54UT), 3 - for 15 Jan 2005 event (22:25UT), 4 - for 17 Jan 2005). (B) - for II HCME kinematics type (where 1 - for 22 Aug 2005 event, 2 - for 23 Aug 2005 event).}
   \label{F-simple}
\end{figure}

  %{\S}{\bf --- two HCME types} \\
We assume that there are two HCME types differing in their time-dependant velocity variations. What may the $V(t)$ profile for the HCME under our consideration depend on? One could assume that the HCME time-dependant velocity variation depends on the ejection source location. Indeed, the four HCMEs under consideration with the same type of $V(t)$ profile originated in the active regions located near the solar disk center, while the two other-type ejections originated in the active regions located near the limb. However, this factor, most likely, is not a determining one in the formation of the HCME velocity $V(t)$ observed profiles. For example, the results from (Temmer et al., 2010), where the $V(t)$ profiles, as in the HCME under our consideration with sources near the solar disk center, were obtained for limb CMEs, validate such a conclusion.

  %{\S}{\bf --- the properties of the active region spot groupings for explanation of these ejections origination} \\
From our standpoint, the properties of the active region spot groupings where these ejections originated are the most likely factor determining the HCME velocity time profiles. Here, we note some distinctions for the active regions where HCMEs with different $V(t)$ profiles originated. Figure 9 exhibits the spot groupings of the active regions where all HCMEs under consideration originated (from the MDI data). One can see clearly that the active regions, where the HCMEs with the velocity profile containing the rapid acceleration to the maximal value, velocity rapid drop, and, finally, velocity slow variation originated, turned out more complicated than the active regions where the HCMEs with the type-II velocity profile formed. The magnetic induction maximal values in the two types of active regions.

\begin{figure}    %%%%%%%%%%%%%%%%%% FIGURE 9
   \centerline{\includegraphics[width=1.0\textwidth,clip=]{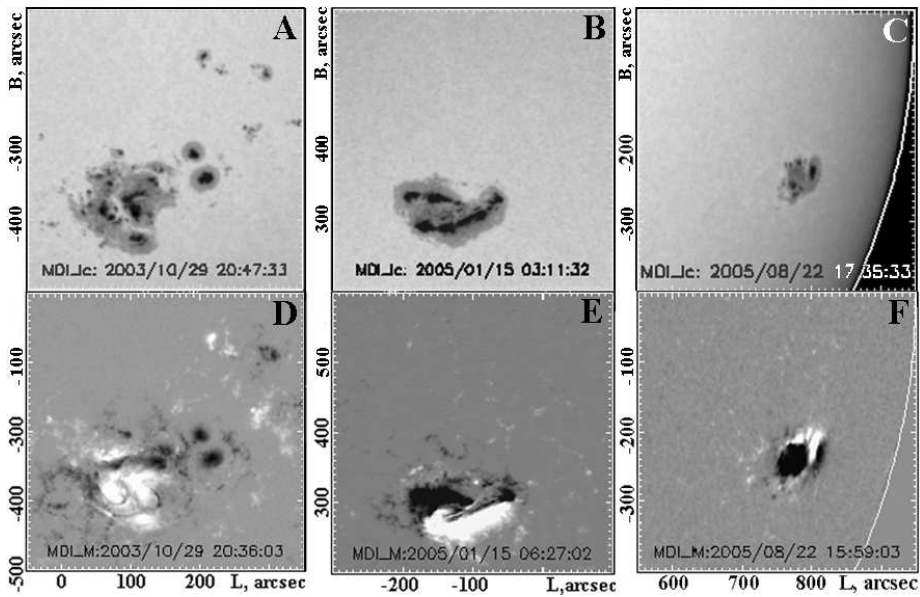}}
   \caption{Groupings of sunspots and the photosphere magnetic field distribution for the two types of HCME. (A-C) – the MDI continuum images of NOAA 10486, 10720, 10798 in which for HCME events on 29 Oct 2003 (A), 15 Jan 2005 (B), 22 Aug 2005 (C), (D-F) – MDI magnetograms of these active regions.}
   \label{F-simple}
\end{figure}

  %{\S}{\bf --- it is the active region properties that determine the kinematics of the HCMEs} \\
We note another fact that indirectly testifies to the fact that it is the active region properties that determine the kinematics of the HCMEs originating in them. For example, the 15 Jan 2005 HCMEs (two ejections were observed that day) and the 17 Jan 2005 HCME originated in the same active region NOAA 10720, but at different instants. However, all of them had the same $V(t)$ variation character. Accordingly, the 22 Aug 2005 and 23 Aug 2005 HCMEs observed in the active region NOAA 10798 at different instants also had similar $V(t)$ profiles differing from the profiles in the previous three events. Thus, we can conclude that the active region properties determine the $V(t)$ profile of a CME originating in it.

\subsection{Comparison of the kinematics of a CME body and associated shock}%%%%%%%%%%%%%%
  \label{S-equations}
When building up the HCME velocity time dependences (Figures 2,6,4 and 8) within the LASCO C2 and C3 fields-of-view, we used the ejections' LE positions from the directory \url{(http://cdaw.gsfc.nasa.gov/CME_list/HALO/halo.html)}. In this directory, data on the locations of the most rapid HCME peculiarity are presented. At the same time, it is not stipulated that, in fact, it might be not the HCME body boundary, but rather a shock front propagating ahead of a rapid coronal mass ejection. Therefore, integrating the HCME velocity values discovered from SXI with the ejection velocity values determined from \url{(http://cdaw.gsfc.nasa.gov/CME_list/HALO/halo.html)} may turn out to be incorrect in cases when we detected not the velocity of the shock wave but the velocity of the HCME body in the SXI field-of-view, while it was the HCME associated shock velocity that was detected from the LASCO data. Detecting a shock from the SXI data entails great difficulties due to the poor quality of the images obtained with this instrument. At the same time, one may reliably identify a shock from the LASCO data in certain cases.

\begin{figure}    %%%%%%%%%%%%%%%%%% FIGURE 10
   \centerline{\includegraphics[width=1.0\textwidth,clip=]{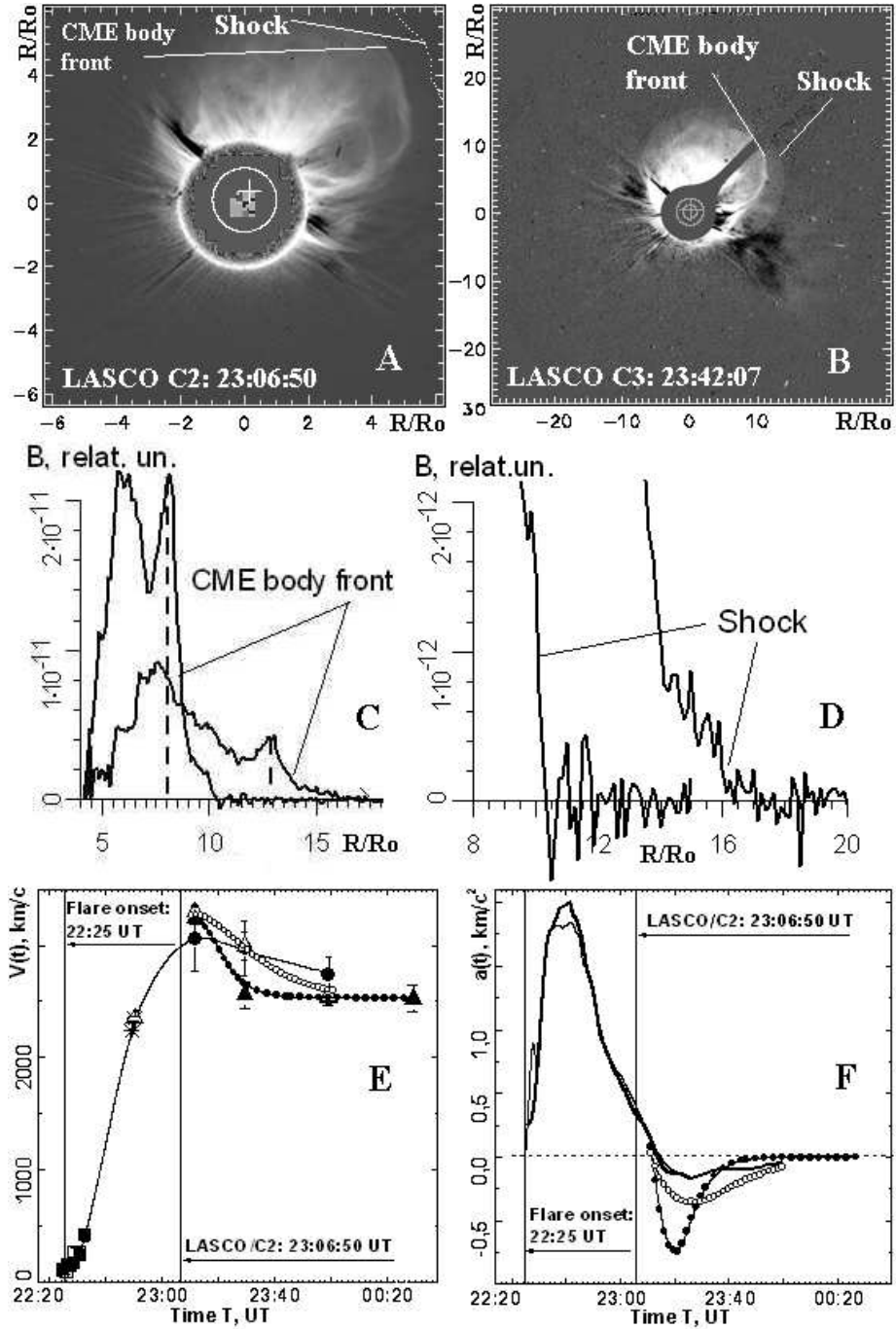}}
   \caption{Comparison of the kinematics of the CME body and the associated shock within the LASCO C2 and C3 fields-of-view for the 15 Jan 2005 event (23:06:50UT). (A),(B) present the running ratio images of the corona with CME. The bright area is the CME body. The diffuse area is presumably shock-compressed plasma behind the shock front. (C) shows the CME front detected when scanning the brightness in the bright area for two instants along the direction indicated in ($http://cdaw.gsfc.nasa.gov/CME_list/$) for 15 Jan 2005 event (23:06:50UT). (D) presents the collisionless shock detected when scanning the brightness in the diffuse area for two instants also. (E) shows the time dependences of the CME body front velocity (open circles) and of the shock velocity (filled-in black circles). Other symbols in Figure 10 (E) correspond to those in figure 2. (F) shows the acceleration $a(t)= dV(t)/dt$ of the CME body (open circles) and of the shock (filled-in black circles). The thin line shows the time dependence of acceleration of the LE of white loop-like structures on running ratio SXI images and the high-speed CME front on running ratio LASCO C2 and C3 images. The thick line presents the time dependence of the acceleration of the LE of dark structure accompanying the white loop-like structures.}
   \label{F-simple}
\end{figure}

  %{\S}{\bf --- how much the CME body V(t) profiles will differ from those for the associated  shock} \\
In this section, we will try to reveal how much the CME body $V(t)$ profiles will differ from those of the associated  shock (using the LASCO data) for one event.

  %{\S}{\bf --- all the HCMEs were associated with shocks} \\
Note that all the HCMEs studied were associated with shocks. This follows indirectly from the brightness distribution character in these ejections. Figure 10(A,B) shows an example for such a distribution within the LASCO C2 and C3 field-of-view for the 15 Jan 2005 (23:06:50 UT) HCME obtained from \url{(http://cdaw.gsfc.nasa.gov/CME_list/daily_movies/)}. The ejection's first appearance within the LASCO C2 field-of-view is indicated in brackets next to the event date. To build up the image in Figure 10(A), we used the moving difference when the image at the nearest earlier instant $t_{\rm 0}$ was deducted from the corona image at instant $t_{\rm 1}$. One can single out the brighter part and the diffuse structure surrounding it in the HCME image. The first is considered the coronal ejection body, and the diffuse structure is assumed to be shock-compressed plasma behind the shock front (Gopalswamy et al., 2010).

  %{\S}{\bf --- detection of the shocks} \\
The shock can be singled out by scanning brightness in the diffuse structure area in the direction perpendicular to the tangent to this structure's boundary. We scanned the brightness along the radial direction from the solar disk center in the point where the diffuse structure boundary was observed at its sharpest. In the brightness distributions we discovered boundary brightness jumps within a $(1-2)\delta R$ spatial scale, where $\delta R$ is the coronagraph's spatial resolution (Figure 10(C,D)). For the LASCO C3, $\delta R\approx 0.125R_{\rm o}$. This gives grounds to consider such jumps collisionless shocks with the front size much smaller than $\delta R$. To single out a shock against random noise brightness variations, we used various methods to reduce the noise level, or, in other words, methods improving the signal-to-noise ratio.

  %{\S}{\bf --- the HCME body front width} \\
Figure 10(C,D) shows brightness distributions within the ejection body along the same direction for three instants. From this Figure, it follows that the HCME body is restricted by a relatively steep front whose amplitude diminished rapidly with distance. But the HCME body front width is manifold bigger than $\delta R$.

  %{\S}{\bf --- the time dependences for the shock velocity and the HCME body} \\
Figure 10(E) presents the time dependences for the shock velocity and the HCME body for the 15 Jan 2005 event (11:06:50 UT) from the LASCO data. The shock velocity was determined as the diffuse area boundary velocity, while the HCME body velocity was found as its mid-front velocity. Figure 10(A) shows the direction along which there were the velocities of the structures specified. One can see that the shock velocity $V_{\rm SH}(t)$ at all the instants when it was possible to measure is more than the HCME body velocity $V_{\rm B}(t)$. The character of the shock velocity variation and the HCME body velocity variation differ significantly. For $\sim 20$ minutes, $V_{\rm SH}(t)$ and $V_{\rm B}(t)$ diminished synchronously, and then the ejection body deceleration became stronger than the shock deceleration. After $\approx 24$ minutes, the HCME body velocity started to vary slowly while the shock acceleration practically did not vary. As a result, at larger distances of $\approx 30R_{\rm o}$ the shock reached the HCME body velocity. Figure 12(F) shows the HCME body and shock accelerations $a(t)=dV(t)/dt$.

  %{\S}{\bf --- we can not separate the HCME shock and HCME body in field-of-view of SXI telescope} \\
Unfortunately, building up individually the aggregated dependences of the HCME body velocity and the associated shock velocity from the SXI and LASCO data does not appear possible yet since, as we noted above, no one has managed to separate the shock from the HCME body within the SXI field-of-view so far.

\subsection{Some summary results}%%%%%%%%%%%%%%
In (Zhang and Dere, 2006) it is shown that there is an inverse dependence between the mean acceleration integrated values $V_{\rm MAX}/t_{\rm ACC}$ and $V_{\rm MAX}/t_{\rm SXR}$ on the one hand, and $t_{\rm ACC}$ and $t_{\rm SXR}$ on the other. Here, $V_{\rm MAX}$ is coronal mass ejection maximal velocity, $t_{\rm ACC}$ is the main acceleration duration (i.e., the time interval from the ejection velocity sharp augment onset till the moment when the CME velocity reaches its maximal value; $t_{\rm SXR}$ is the soft X-ray intensity ($I_{\rm SXR}$) rise time from the flare onset instant till when $I_{\rm SXR}$ becomes maximal. Figure 11 shows the regression dotted line for this dependence from (Zhang and Dere, 2006). This Figure also exhibits two more correlations. Solid (open) circles present the $V_{\rm MAX}/t_{\rm ACC}$ ($V_{\rm MAX}/t_{\rm SXR}$) dependence on $t_{\rm ACC}$ ($t_{\rm SXR}$) for the events under under consideration. The solid line represents the regression line for this point set. Figure 11 also presents the correlation between the HCME a(t) maximum acceleration $a_{\rm MAX}$ and $t_{\rm ACC}$ ($t_{\rm SXR}$) (solid and open triangles) for the HCMEs considered in this study. Points show the regression line corresponding to this dependence. Both dependences on $t_{\rm ACC}$ ($t_{\rm SXR}$) also testify to the existence of an inverse correlation between the mean (maximal) acceleration and the main acceleration duration or the soft X-ray intensity rise duration. This is consistent with the results from (Zhang and Dere, 2006). At the same time, we note that the dependences we obtained turned out higher than the dependence from (Zhang and Dere, 2006). This is associated with that the HCMEs studied here were characterized, on an average, by the acceleration greater values than in (Zhang and Dere, 2006). Since we observe the HCME velocity projections on the sky plane, real 3-D accelerations may possess still greater values.

\begin{figure} %%%%%%%%%%%%%%%%%% FIGURE 11
\centerline{\includegraphics[width=1.0\textwidth,clip=]{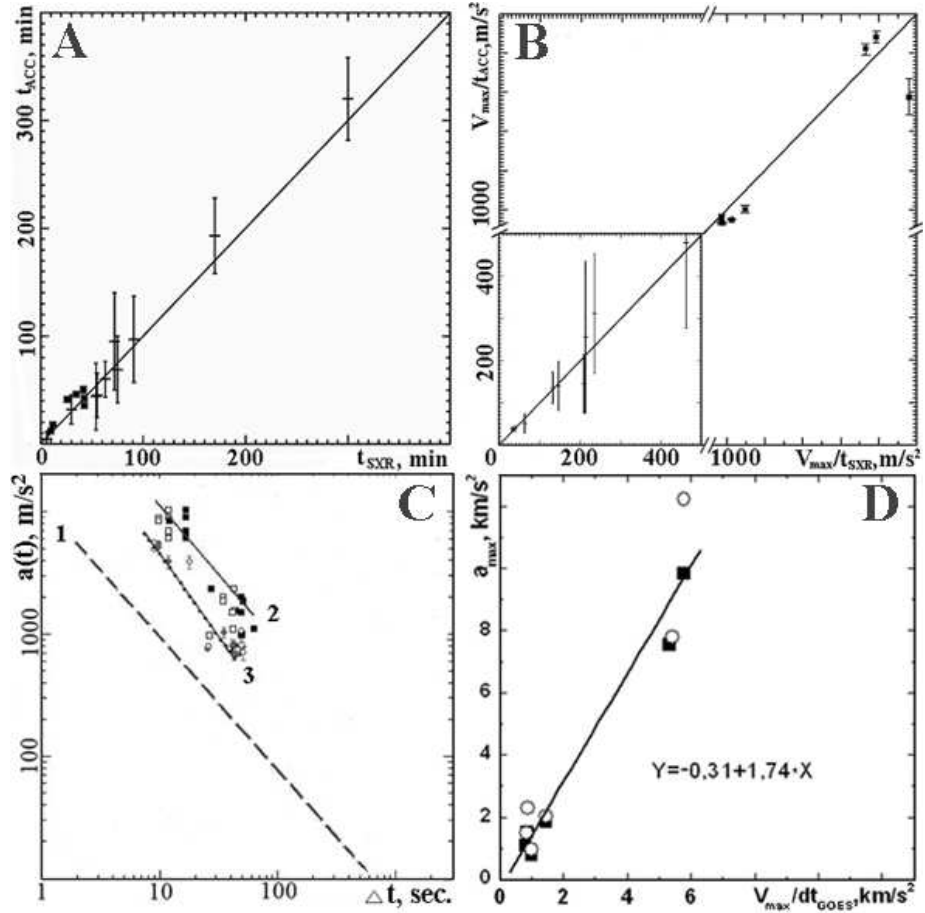}}
\caption{The association of the HCME acceleration with the duration of the soft-X ray intensity rise from the associated flare area. Dashed line (1) is the regression line from (Zhang and Dere, 2006) for the mean acceleration dependence $V_{\rm MAX}/t_{\rm ACC}$ and $V_{\rm MAX}/t_{\rm SXR}$ on $t_{\rm ACC}$ on the one hand, and on $t_{\rm SXR}$ on the other. Here, $V_{\rm MAX}$ is the HCME maximal velocity, $t_{\rm ACC}$ is the measured duration of the main ejection acceleration; $t_{\rm SXR}$ is the rise time $I_{\rm SXR}$ from the flare onset to the $I_{\rm SXR}$ maximum. Solid and open squares reflect the correlation between $a(t)$ acceleration maximum $a_{\rm MAX}$ and $t_{\rm ACC}$ $(t_{\rm SXR})$ for the HCMEs under consideration.The solid line is the regression line for this point set. Solid (open) circles (3) show the dependence $V_{\rm MAX}/t_{\rm ACC}$ $(V_{\rm MAX}/t_{\rm SXR})$ on $t_{\rm ACC}$ $(t_{\rm SXR})$ for the events under our consideration.The line with dots show the regression line for this dependence.}
\label{F-simple}
\end{figure}

  %{\S}{\bf --- the CME trajectories in the picture plane} \\
Figure 12(A) shows how the HCME motion direction $PA_{\rm CME}$ within the SXI field-of-view at a fixed instant is determined, and Figure 12(B) presents the CME trajectories in the picture plane ($PA_{\rm CME}(t)$) from the SXI and LASCO data. One can see that as the HCME translational velocity augmented, the ejections' trajectories either turned out rectilinear or, for the HCMEs observed in the northern hemisphere, they deviated towards the North Pole, and for the HCMEs observed in the southern hemisphere they deviated towards the South Pole.

\begin{figure}    %%%%%%%%%%%%%%%%%% FIGURE 12
   \centerline{\includegraphics[width=1.0\textwidth,clip=]{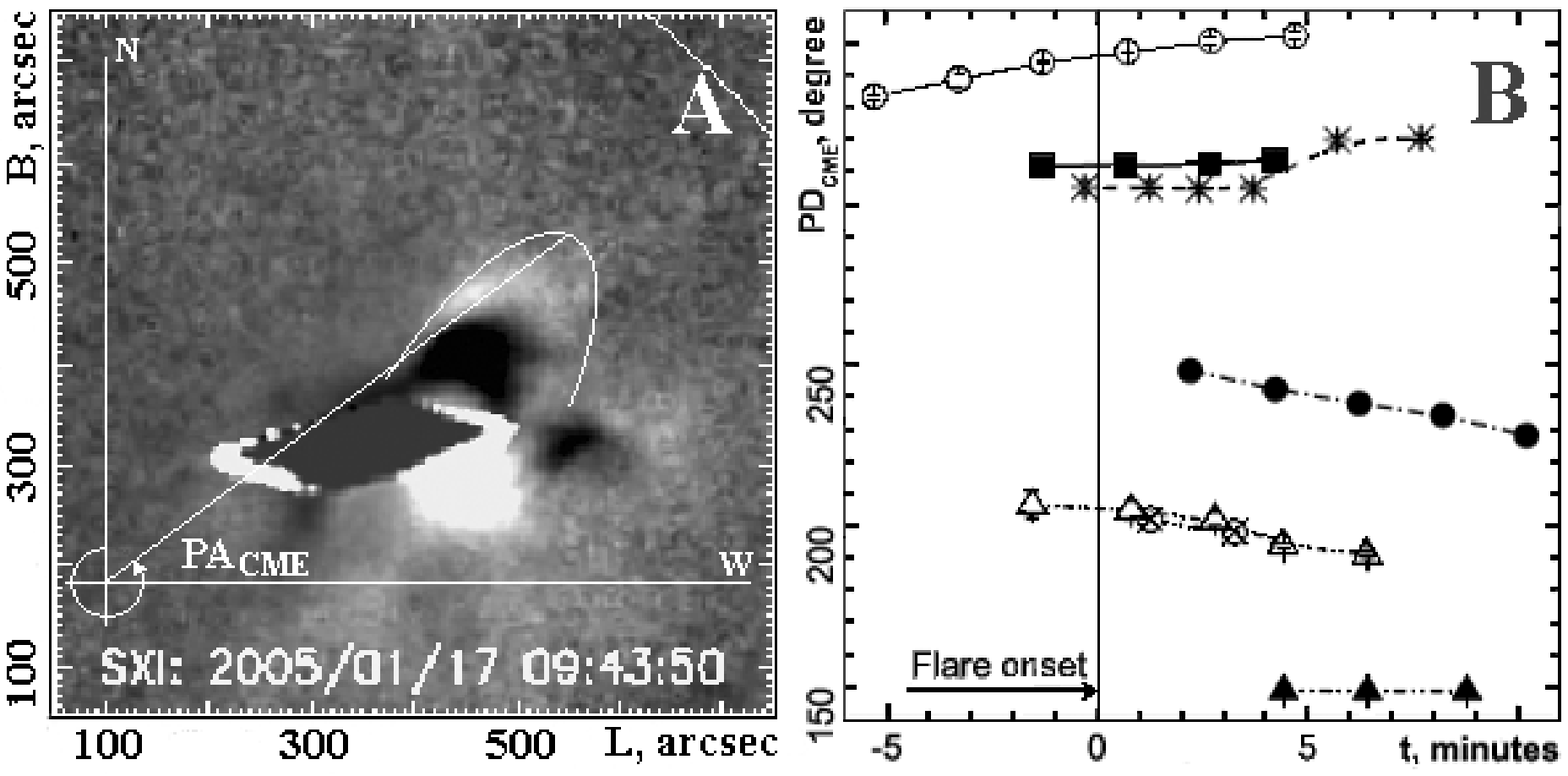}}
   \caption{(A) presents the HCME motion direction $PA_{\rm CME}$ within the SXI field-of-view at a fixed instant. (B) shows the HCME trajectories in the picture plane ($PA_{\rm CME}(t)$) from the SXI and LASCO data; where filled-in black triangles mark $PA_{\rm CME}$ values for the 29 Oct 2003 HCME event; circles mark the 15 Jan 2005 HCME event (05:54UT); double crosses mark the 15 Jan 2005 HCME event (22:25UT); squares mark the 17 Jan 2005 HCME event, crossing circles mark the 22 Aug 2005 HCME event, filled-in black circles mark the 23 Aug 2005 HCME event and triangles mark one loop-like structure (seen on 29 Oct 2003 in NOAA 10486), marked on Figure 6(A) by $Fw_{\rm 1}$, seen a moment before the loop-like structures became one wide loop-like structure that looked like to a flux-rope.}
   \label{F-simple}
\end{figure}

  %{\S}{\bf --- the HCME height-to-width ratio} \\
Figure 13(A) exhibits how the coronal mass ejection height $d_{\rm H}$ and width $d_{\rm W}$ are determined, and Figure 13(B) presents the time variation of the HCME height-to-width ratio ($d_{\rm H}/d_{\rm W}$) within the SXI field-of-view. We revealed a unique feature of the time variation for this parameter: for all the HCMEs under consideration the ejection height at the initial stage of its motion augmented mor rapidly than its width did. After reaching the maximal value, the variation character for $(d_{\rm H}/d_{\rm W})(t)$ with various ejections are found to be different. For three HCMEs after reaching maximum, this parameter diminished dramatically with time, i.e., there came a short period when the HCME varied crosswise more than in its propagation direction.

\begin{figure}    %%%%%%%%%%%%%%%%%% FIGURE 13
   \centerline{\includegraphics[width=1.0\textwidth,clip=]{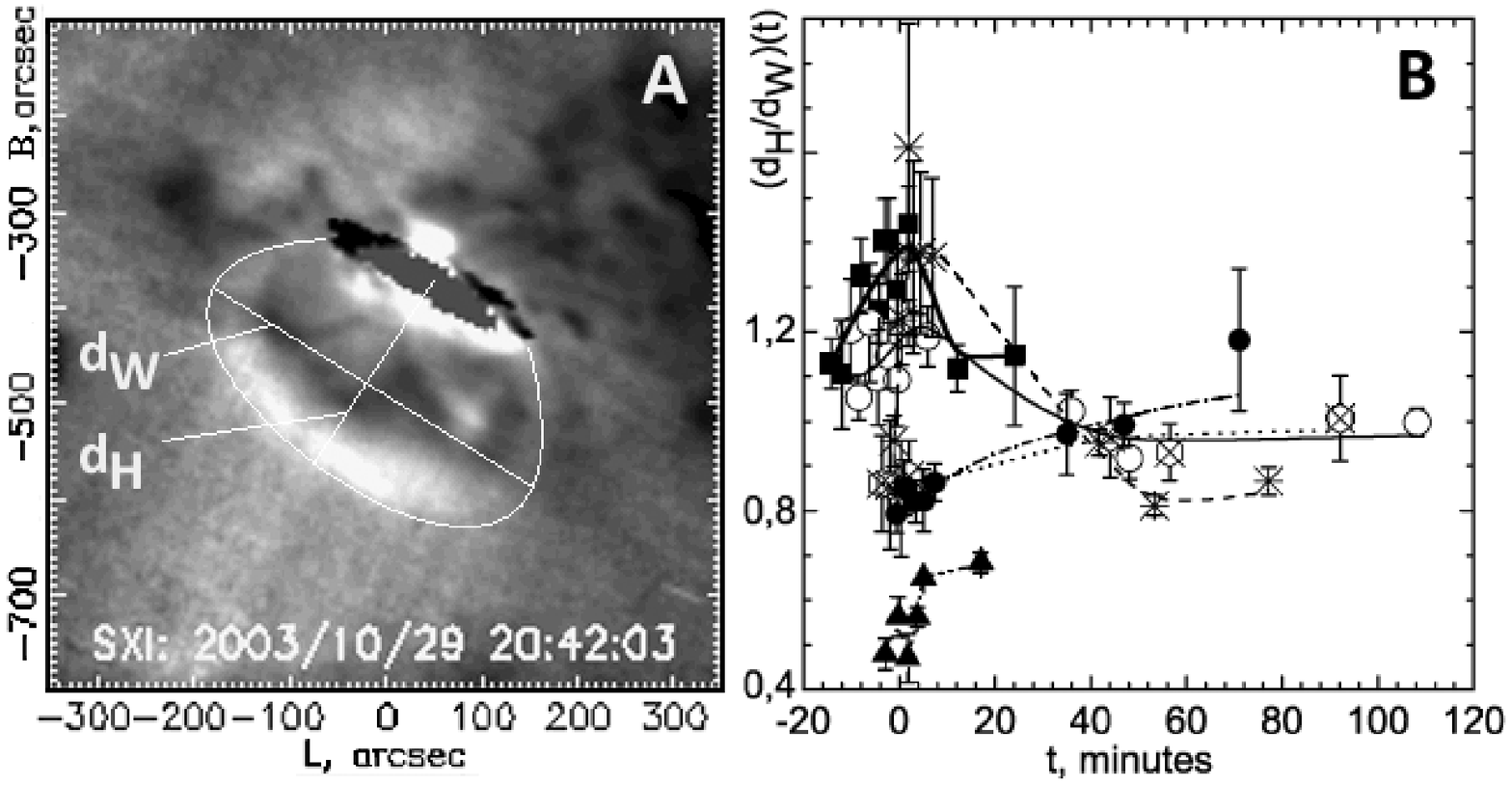}}
   \caption{(A) presents the way the CME height $d_{\rm H}$ and width $d_{\rm W}$ were determined. (B) is the $d_{\rm H}/d_{\rm W}$ ratio time variation within the SXI field-of-view. Symbols correspond to the symbols in the figure 12(B).}
   \label{F-simple}
\end{figure}

  %{\S}{\bf --- the angular size variation for the HCMEs} \\
Finally, Figure 14(B) illustrates the peculiarities of the angular size variation for the HCMEs under consideration within the SXI field-of-view. The angular point determining the ejection size is placed into the solar disk center, and the rays forming the angle, adjoin the ejection boundary extreme points (see Figure 14(A)). As one can see from this Figure, the angular size $2\alpha$ for all the HCMEs under consideration augments with time within the SXI field-of-view, but the character of $2\alpha(t)$ variation changes ejection-to-ejection. The angular size $2\alpha$ increases to about 4.5 times its initial value. And the speed of change of the angular size $2\alpha$ varies within a 5-10 minute range, where the parameter $t$ is the period of time over which the angular size increased 1.5 times its initial value.

\begin{figure}    %%%%%%%%%%%%%%%%%% FIGURE 14
   \centerline{\includegraphics[width=1.0\textwidth,clip=]{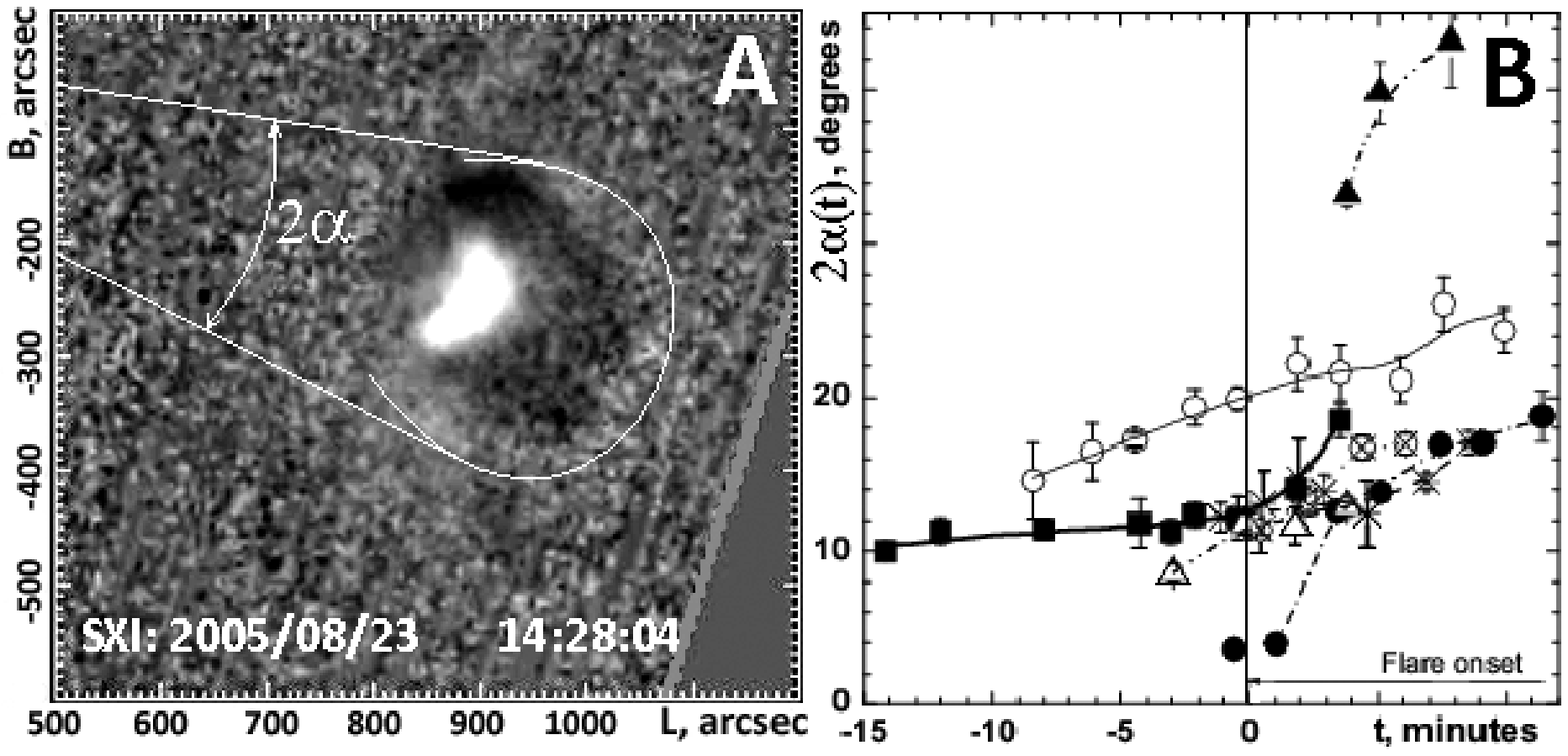}}
   \caption{(A) presents the way the HCME angular size $2\alpha$ was determined. (B) is the change in $2\alpha (t)$ with time. Symbols correspond to those in figure 12(B).}
   \label{F-simple}
\end{figure}

  %{\S}{\bf --- HCME frontage values} \\
In Figure 15 the change of ejection front width for all considered ГКВМ is shown. The ejection front width was determined as the distance between the top and base of the front according to distribution of brightness along axis ГКВМ (see Figure 1(D,E)). It is seen that the front width of all the coronal mass ejections analyzed increases variously in course of time. At the same time the average speed of increase of ejection front width varies considerably from event to event. Ejection front width starts to vary most quickly after the flare beginning.

\begin{figure}    %%%%%%%%%%%%%%%%%% FIGURE 15
   \centerline{\includegraphics[width=1.0\textwidth,clip=]{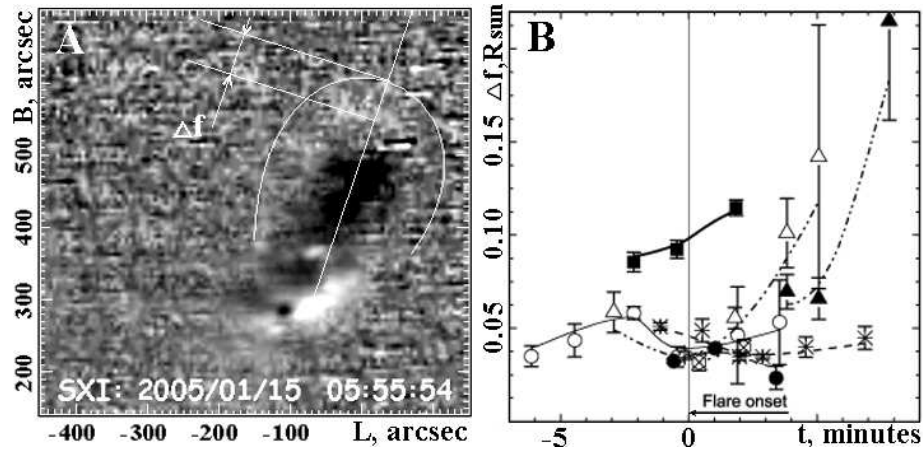}}
   \caption{(A) presents an example of the way the ejection front width of the loop-like structure  of the 15 Jan 2005 HCME event (05:54UT) was determined, (B) is the change in the ejection front width with time for all considered ГКВМ. Symbols correspond to those in figure 12(B).}
   \label{F-simple}
\end{figure}

\section{Discussion and conclusion} %%%%%%%%%%%%%%%%%%%%%%%%%%%%%%%%%%%%%%%%
      \label{S-Conclusion}
We investigated the laws of the initial stage of movement of six halo coronal mass ejections using the data of several telescopes, and, first of all, GOES-12/SXI, SOHO/LASCO, SOHO/EIT. For the analysis, these HCMEs were selected from the group of the fastest ejections (with $V>1500~km/s$) connected with powerful flares. The choice of such events was determined, first, by the role of fast HCME in space weather, and second, the opportunity to reveal them at the stage of their formation. The HCME events selected for analysis took place place in 2003 and 2005. We used SXI as the telescope with high time cadence, though during the period when the HCME events analysed took place the telescope TRACE with higher time cadence was working. But the use of TRACE images for studying the initial stage of the movement of all selected ejections turned out to be impossible because of the limited field of view of this telescope and some features of its functioning.

We were not able to find eruptive filaments connected with ejection in any of the HCME considered on the $H_{\rm \alpha}$ images. At the same time, we established that, before the occurrence of ejection in five of the six HCME events (except the HCME event on 29 Oct 2003), the loop-like structure identified as the beginning of HCME using SXI data, is observed as a single emission coronal loop in the extreme ultraviolet spectral line at $\lambda =19.5~nm$ on the EIT data image. All the loop-like structures become more active and begin slow forward movement before the beginning of the flare connected with the HCME. In this work we do not discuss questions concerning the reasons for formation and activation of these emission loop-like structures. It is possible that it is the flux-ropes existing in the active area for some time and becoming more active as a result of the infringement of magnetic field balance. Possible mechanisms of such balance infringement are discussed in several works [Forbes et al., 2006; Howard, 2011]. The occurrence of an HCME event on 29 Oct 2003 was apparently connected with the movement not of a separate loop, as in the other events studied, but with the movement of an arcade of loops or groups of unconnected loops.

We have shown that the time dependence of the $V(t)$ speed profile of the fast HCME can be of two types. So, for the HCME event on 17 Jan 2005, the $V(t)$ profile quickly reaches maximum value before entering the field of view of LASCO C2, then it falls sharply over a short period of time and then slowly decreases (see Figure 2А). The two HCME observed on 15 Jan 2005 in this active region have close kinematic properties. The HCMEs consistently observed on 22 and 23 August 2005 close to the limb from the same active region (see Figure 4А) have another $V(t)$ profile change character. For these HCME events the $V(t)$ profile quickly increases before the appearance of the ejection into the field of view of LASCO C2 coronagraph and then it continues to increase slowly and reaches maximum value at a large distance from the Sun's surface. It turned out that the HCMEs arising in one and the same active region have identical time speed profiles. We  assume that HCMEs with such different time speed profiles relate to two different classes of coronal mass ejections. It was ascertained that HCMEs of the first type are formed in active regions with a complex configuration of sunspots and magnetic field and with a large sunspot area, and HCMEs of the second class are formed in active regions with more simple sunspot and magnetic field configurations.

As in a number of previous papers by other authors we came to the conclusion that the kinematics of the HCMEs analysed is synchronized with the time variation of intensity of soft X-ray radiation $I_{\rm SXR}(t)$ from the flare area connected with the ejection [Gallagher et al., 2003; Zhang and Dere, 2006; Mari\v{c}i\'{c} et al., 2007; Patsourakos et al., 2010]. It turned out, that the $t_{\rm ACC}$ time of the basic acceleration of all HCMEs investigated is close to the time of increase $I_{\rm SXR}(t)$ from the flare's beginning till the moment of $I_{\rm SXR}$ maximum value. The HCME maximal measured acceleration adopts the value $a_{\rm MAX}\approx 0.9-10~km/s^2$ and is close to the value obtained by dividing the HCME maximal velocity $V_{\rm MAX}$ by the X-ray intensity rise time $t_{\rm SXR}$. It is shown that there is an inverse correlation between $V_{\rm MAX}/t_{\rm ACC}$ and $V_{\rm MAX}/t_{\rm SXR}$ on the one hand and $t_{\rm ACC}$ and $t_{\rm SXR}$ - on the other. This is consistent with the results in [Zhang and Dere, 2006]. At the same time, this dependence in our study was obtained for the values $V_{\rm MAX}$ and $а_{\rm MAX}$ that are much higher than in [Zhang and Dere, 2006]. Acceleration $a(t)$ of HCMEs is closely connected with the intensity of rigid X-ray radiation $I_{\rm HX}(t)$ from the flare area. But conformity between $a(t)$ and $I_{\rm HX}(t)$ values varies from event to event.

At the initial movement stage, the trajectories of several HCMEs investigated are curvilinear and deviate from the equator. At the same time, the HCMEs originating in the Sun's northern hemisphere deviate towards the North Pole, whereas the HCMEs originating in the southern hemisphere deviate towards the South Pole.

We studied the time variation in the HCME angular size. The angular size of all the HCMEs under consideration is shown to augment with time from their translational motion onset, and may increase by a factor of $\sim 4.5$ within the SXI field-of-view. The characteristic time for the increase in angular size by a factor of 1.5 is 5-10 minutes. The initial angular size of four HCMEs on SXI data images does not exceed $10^\circ$, and for one ejection - $14^\circ$. For the HCME event on 29 Oct 2003 there is uncertainty as to the estimation of the initial angular size of the ejection since it is formed from three small loops or from a loop arcade. After the ejection was formed its initial angular size was about $32^\circ$.

The change  in the time variation ratio $d_{\rm H}/d_{\rm W}(t)$ of the HCME longitudinal size and its cross-section size was investigated. We showed that for each HCME under consideration its longitude-to-cross size ratio increases with time within the first minutes of motion, and then the character of change in this parameter turns out different for each ejection depending on time. In some cases the longitudinal size of the HCME changes quicker than the crosswise, in others vice versa. This ratio turns out to be close to unity within the LASCO field-of-view for three HCMEs, and does not vary for some time. This may reflect the HCME transition to the self-similar expansion mode. But taking into account that we are investigating the movement of halo coronal mass ejections such an inter-pretation $d_{\rm H}/d_{\rm W}(t)$ in the field of view of LASCO coronagraphs can appear incorrect.

We also investigated the time variation of HCME's front width $\Delta f(t)$. The definition of the HCME's front width is shown in Figure 1(E) where the HCME's front is limited by the top and the bottom plusses (and see also the illustration on Figure 15(A)). It has been established, that the HCME's front width for five of the six events of HCMEs studied increases non-monotonously with time. In one case the tendency to increase of $\Delta f(t)$ arises at the end of the time period for which it was possible to measure this parameter (see Figure 15).

%%%%%%%%%%%%%%%%%%%%%%%%%%%%%%%%%%%%%%%%%%%%%%%%%%%%%%%%%%%%%%%%%%%%%%%%%%%
\begin{acks}
We established that the front width for all the HCMEs under consideration augments non-monotonously with time in general case.
The authors are grateful to the GOES and GOES-12/SXI, RHESSI, TRACE, SOHO/EIT, SOHO/LASCO and MarkIV (MLSO) teams for the possibility to freely use the data from these instruments. The authors appreciate V.V. Grechnev's most beneficial discussions of this paper.
\end{acks}

%%% BIBLIOGRAPHY %%%%%%%%%%%%%%%%%%%%%%%%%%%%%%%%%%%%%%%%%%%%%%%%%%%%%%%%%%%

\end{article}
\end{document}